\documentclass[numberedappendix]{emulateapj}

\usepackage{epsfig, natbib}
\tightenlines

\countdef\decade=200
\advance\decade by \year
\countdef\hours=201	
\advance\hours by \time
\divide\hours by 60
\countdef\mins=202
\advance\mins by \hours
\multiply\mins by 60
\multiply\hours by 100
\countdef\miltime=203
\advance\miltime by \hours
\advance\miltime by \time
\advance\miltime by -\mins

\def\mathbi#1{\textbf{\em #1}}

\newcommand{\gtsim}{\protect\raisebox{-0.5ex}{$\:\stackrel{\textstyle >}
        {\sim}\:$}}

\newcommand{\msun}{M_{\odot}}

\newcommand{\vecv}{\mathbi{v}}
\newcommand{\vecx}{\mathbi{x}}
\newcommand{\vecp}{\mathbi{p}}

\usepackage{color}

\begin{document}

\title{Radiation-Hydrodynamic Simulations of the Formation of Orion-Like Star Clusters\\
I. Implications for the Origin of the Initial Mass Function}

\slugcomment{Submitted to the Astrophysical Journal}

\shorttitle{Radiation-Hydrodynamic Simulations of Cluster Formation}
\shortauthors{Krumholz et al.}

\author{
        Mark R. Krumholz\altaffilmark{1},
        Richard I. Klein\altaffilmark{2, 3}, and
        Christopher F. McKee\altaffilmark{3,4}}

\altaffiltext{1}{Department of Astronomy and Astrophysics,
         University of California, Santa Cruz, CA 95064;
         krumholz@ucolick.org}
\altaffiltext{2}{Lawrence Livermore National Laboratory, P.O. Box 808, L-23, Livermore, CA 94550}
\altaffiltext{3}{Department of Astronomy and Astrophysics, University of California, Berkeley,
Berkeley, CA 94720}
\altaffiltext{4}{Department of Physics, University of California, Berkeley,
Berkeley, CA 94720}

\begin{abstract}
One model for the origin of typical galactic star clusters such as the Orion Nebula Cluster (ONC) is that they form via the rapid, efficient collapse of a bound gas clump within a larger, gravitationally-unbound giant molecular cloud. However, simulations in support of this scenario have thus far have not included the radiation feedback produced by the stars; radiative simulations have been limited to significantly smaller or lower density regions. Here we use the ORION adaptive mesh refinement code to conduct the first ever radiation-hydrodynamic simulations of the global collapse scenario for the formation of an ONC-like cluster. We show that radiative feedback has a dramatic effect on the evolution: once the first $\sim 10-20\%$ of the gas mass is incorporated into stars, their radiative feedback raises the gas temperature high enough to suppress any further fragmentation. However, gas continues to accrete onto existing stars, and, as a result, the stellar mass distribution becomes increasingly top-heavy, eventually rendering it incompatible with the observed IMF. Systematic variation in the location of the IMF peak as star formation proceeds is incompatible with the observed invariance of the IMF between star clusters, unless some unknown mechanism synchronizes the IMFs in different clusters by ensuring that star formation is always truncated when the IMF peak reaches a particular value. We therefore conclude that the global collapse scenario, at least in its simplest form, is not compatible with the observed stellar IMF. We speculate that processes that slow down star formation, and thus reduce the accretion luminosity, may be able to resolve the problem.
\end{abstract}

\keywords{ISM: clouds --- radiative transfer --- stars: formation --- stars: luminosity function, mass function --- turbulence}

\section{Introduction}
\label{sec:intro}

The origin of the stellar initial mass function (IMF) is one of the outstanding problems in the modern theory of star formation. While there have been numerous analytic and numerical studies purporting to explain its origin (e.g.\ see the review by \citealt{mckee07a}, and references therein), much of this work has been hampered by the limited number of physical processes that are included in models of how gas fragments. In particular, while both simulations and analytic work reveal that how gas fragments into stars is extremely sensitive to how the temperature of the gas varies with its density \citep{larson05a, jappsen05a}, it has been common until very recently to approximate this relationship with a simple equation of state \citep[e.g.][]{bate05a, bonnell06d, offner08a,hennebelle11a}. Since the characteristic masses of the stars formed in a collapse are largely determined by the temperature-density relationship, predictions about the location of the IMF peak in these simulations are only as good as their adopted equations of state.

Given this realization, attention in recent years has shifted to models that attempt to determine the temperature-density relationship from first principles, or to include a self-consistent treatment of the thermal evolution of the gas in numerical simulations. In the former category, much work has focused on the effects of imperfect coupling between gas and dust grains on gas thermodynamics. For example, \citet{larson05a} and \citet{elmegreen08a} both argue that the characteristic stellar mass is set by the Jeans mass at the density and temperature where dust grains and gas become thermally coupled due to collisions. According to these models, at low densities where grain-gas coupling is poor, the gas is slightly sub-isothermal, while at higher densities it is slightly super-isothermal, and this effect favors fragmentation near the coupling density.

However, this argument faces a major difficulty in explaining the IMF in the dense, cluster-forming regions where much Galactic star formation appears to take place. The density at which grains and gas become well-coupled is $\sim 10^4 - 10^5$ H$_2$ molecules cm$^{-3}$ \citep{goldsmith01a}, roughly independent of the metallicity and of ambient radiation field intensity \citep{krumholz08a, elmegreen08a, krumholz11b}. In comparison, observations now show that the typical site of star cluster formation has a mass of $\sim 10^3-10^4$ $\msun$, and a radius of $\sim 0.3-0.5$ pc (e.g.\ see \citealt{shirley03a}, \citealt{faundez04a}, \citealt{fontani05a}, or the summary plot combining these data sets in \citealt{fall10a}), giving a mean density $\sim 10^5$ cm$^{-3}$. Similarly, the present-day Orion Nebula Cluster has a mass of $2400$ $\msun$ within a half mass radius of $0.8$ pc, corresponding to $2\times 10^4$ cm$^{-3}$, and within the $\sim 0.2$ pc core the mean density reaches $4\times 10^5$ cm$^{-3}$ \citep{hillenbrand98a}. Since the star formation efficiency was certainly less than unity, and the cluster has likely expanded some since the gas was expelled \citep{kroupa01b, tan06a}, the density at which most of the stars formed must have been higher by at least a factor of a few. Thus the typical site of star cluster formation in the Galaxy, of which the ONC is an example, is in the regime where essentially all the mass is at densities where grain-gas coupling is very strong. It is therefore hard to see how grain-gas coupling could be relevant for determining how this gas fragments. This argument can be made even stronger by noting that globular clusters with mean densities $\sim 10^7$ cm$^{-3}$ in their centers, $2-3$ orders of magnitude above the grain-gas decoupling density, also appear to have the same IMF peak as the Galactic field \citep{marchesini09a}.

A second class of models for the temperature-density relationship focuses on the interaction of gas with the radiation produced by stars in the star formation process. In these models one assumes good grain-gas coupling, as is appropriate at the high densities where most stars form. The gas temperature and its relationship with the density is then determined primarily by the light produced by stars in the process of formation. Conceptually, the idea is that the luminosity from an accreting star warms the gas in its immediate vicinity, inhibiting the ability of that gas to fragment, and that this process determines characteristic stellar masses. Analytically, \citet{krumholz06b} and \citet{krumholz08a} have argued that this process explains how massive stars are able to form under certain circumstances, while \citet{bate09a} argues that it can explain the characteristic peak of the IMF.

However, numerical studies of the second class of models have thus far been limited in various ways. \citet{krumholz07a, krumholz10a} and \citet{myers11a} conduct simulations including stellar feedback and radiative transfer (including re-radiation by dust grains, which is the critical process in determining the gas temperature), but focus on single massive cores that do not (and are not expected to) form a full IMF. \citet{commercon10a} report similar simulations focusing on single low-mass cores. \citet{bate09a}, \citet{offner09a}, \citet{price09a}, and \citet{peters10b, peters11a} simulate the formation of star clusters, but consider only low-density regions similar to those found in nearby clouds like Taurus, rather than conditions typical of Galactic star formation sites.\footnote{Although \citet{peters10b, peters11a} study regions with enough mass to form massive stars, the column densities of the regions they simulate are $\sim 0.01$ g cm$^{-2}$, rather than the $\sim 1$ g cm$^{-2}$ typical of Galactic star-forming sites. Their simulated clouds are therefore optically thin even in the near-IR, rendering radiative effects fairly unimportant. In this way their work is closer to that of \citeauthor{bate09a}, \citeauthor{offner09a}, and \citeauthor{price09a} than to the simulations we present here.} As we will see, this makes a large difference in the outcome, because under low-density conditions the regions of heating around each star are non-overlapping, while in denser conditions they are not. Moreover, of these simulations, only \citeauthor{offner09a} and \citeauthor{peters10b} include stellar luminosity, so the amount of heating in the other two simulations is underestimated.

Other simulations of star cluster formation do not include radiative transfer at all, and instead approximate it in various ways. For example, \citet{smith09a} and \citet{urban10a} study the fragmentation of dense gas clouds similar to typical star-forming regions, but they determine the gas temperature around each star via a rough fitting formula based on static radiative transfer calculations. This approximation may be reasonable as long as the heating at a given point is dominated by a single star, but it almost certainly fails once the regions of heating around stars begin to overlap, as occurs in dense regions. 

In summary, to date there have been no simulations capable of studying how the peak of the IMF under typical Galactic conditions, including the all-important effects of stellar feedback and re-radiation by dust grains. The goal of this paper, the first in a series, is to remedy that lack. We use the ORION adaptive mesh refinement radiation-hydrodynamics code to simulate a typical galactic star-forming clump including stellar feedback and radiative transfer. As this is a first attack on the problem, we choose the simplest possible scenario. We do not include magnetic fields or any form of feedback other than radiation, and we allow the initial turbulence in the cloud to decay freely, leading to a rapid global collapse. Our simulation therefore represents a minimalist scenario for the formation of a star cluster such as the ONC similar to that proposed by, for example \citet{bonnell03a}. Previous authors who have studied such conditions report that they produce stellar mass distributions consistent with the observed IMF at all times in the simulation, but it is clear in retrospect that this result simply reflects the imposed equation of state. Our work therefore revisits the critical question of whether such a scenario is capable of reproducing the observed IMF.

The remainder of this paper proceeds as follows. In Section \ref{sec:method} we describe our numerical method and simulation setup. In Section \ref{sec:results} we report the results of our simulations. In Section \ref{sec:discussion} we discuss the implications of our findings, and present simple analytic models to aid in understanding them. Finally, we summarize in Section \ref{sec:summary}.

\section{Simulation Description}
\label{sec:method}

\subsection{Simulation Initial Conditions}
\label{sec:ic}

\begin{deluxetable*}{cccccccccccc}
\tablecaption{Simulations \label{runs}}
\tablehead{
\colhead{Name} &
\colhead{RT?} &
\colhead{$M_c$ ($\msun$)} &
\colhead{$\Sigma_c$ (g cm$^{-2}$)} &
\colhead{$\overline{\rho}$ (g cm$^{-3}$)} &
\colhead{$t_{\rm ff}$ (kyr)} &
\colhead{$L_{\rm box}$ (pc)} &
\colhead{$N_0$} &
\colhead{$L$} &
\colhead{$\Delta x_L$ (AU)} &
\colhead{$t_{\rm fin}/t_{\rm ff}$} &
\colhead{$M_{*,\rm fin}/M_c$}
}
\startdata
LR & Yes & 1000 & 1.0 & $9.4\times 10^{-19}$ & 68.6 & 1.9 & 256 & 4 & 98 & 0.94 & 0.51 \\
HR & Yes & 1000 & 1.0 & $9.4\times 10^{-19}$ & 68.6 & 1.9 & 256 & 5 & 49 & 0.94 & 0.52 \\
ISO & No & 1000 & 1.0 & $9.4\times 10^{-19}$ & 68.6 & 1.9 & 256
& 5 & 49 & 0.94 & 0.65 \\
\enddata
\tablecomments{Col.\ 2: radiative transfer included? Col.\ 3: cloud mass. Col.\ 4: cloud surface density. Col.\ 5: mean volume density. Col.\ 6: mean-density free-fall time. Col.\ 7: linear size of computational domain. Col.\ 8: number of grid cells per linear dimension on the coarsest AMR level. Col.\ 9: maximum AMR level. Col.\ 10: linear cell size on the maximum AMR level. Col.\ 11: time relative to free-fall time to which simulation is evolved. Col.\ 12: total mass of stars at the final evolution time, normalized to the initial cloud mass.
}
\end{deluxetable*}

We conduct two simulations that are identical in every respect except that they have different maximum AMR levels, meaning that the peak resolution is different in the two runs. We refer to these as the low-resolution (LR) and high-resolution (HR) simulations. The two simulations enable us to determine to what extent our results are converged, although we caution that the two simulations differ only in their peak resolution, which is deployed near stars and in regions of high density or large radiation gradients (see Section \ref{sec:conditions}). Thus we have not tested the sensitivity of our results to variations in the resolution used in low density regions far from stars. We also carry out a third simulation with identical initial conditions at the same resolution as run HR, but with an isothermal equation of state, i.e.~with the radiative transfer module in our code disabled. We refer to this as run ISO. This simulation enables us to determine what effects in our simulation are due to radiative transfer.

We summarize the key parameters of the runs in Table \ref{runs}. The initial conditions for both consist of a $M_c = 1000$ $\msun$ spherical gas cloud with a mean surface density $\Sigma_c = 1$ g cm$^{-2}$, corresponding to a mean volume density of $9.4\times 10^{-19}$ g cm$^{-3}$, or $2.4\times 10^5$ H$_2$ molecules cm$^{-3}$. The corresponding cloud radius is  $R_c = \sqrt{M_c / (\pi \Sigma_c)}=0.26$ pc, and we place the cloud in a cubical computational domain of size $L_{\rm box} = 1.9$ pc, roughly four times the cloud diameter. We have chosen this mass and surface density because they are typical of regions of clustered star formation in the Galaxy (e.g.\ see \citealt{shirley03a}, \citealt{faundez04a}, \citealt{fontani05a}, and a summary of the data in Figure 1 of \citealt{fall10a}.). They are also roughly the estimated parameters of the progenitor of the core of the Orion Nebula Cluster \citep[e.g.][]{kroupa01b, tan06a}. It is worth noting that our initial conditions are significantly denser than has been used for some previous simulations of massive star formation. For example, \citet{bonnell03a} use initial mean volume and column densities of $1.3\times 10^{-19}$ g cm$^{-3}$ ($3.3\times 10^4$ cm$^{-3}$) and $0.26$ g cm$^{-2}$, respectively; \citet{peters10b} use $3.9\times 10^{-21}$ g cm$^{-3}$ ($1.0\times 10^3$ cm$^{-3}$) and $0.026$ g cm$^{-2}$. However, our parameter choices are much closer to what is actually observed in regions of massive star formation. For example, in their survey of $146$ Southern massive star-forming regions, \citet{faundez04a} find a typical mass and radius of $5000$ $\msun$ and $0.4$ pc, corresponding to a volume density of $1.2\times 10^{-18}$ g cm$^{-3}$ ($3.1\times 10^5$ cm$^{-3}$) and a column density of $2.1$ g cm$^{-2}$, similar to what we use.

Our initial cloud has a density structure described by
\begin{equation}
\rho = \left\{
\begin{array}{ll}
\rho_c, & r < R_c/2 \\
\rho_c (2r/R_c)^{-1.5}, \qquad & R_c/2 \leq r < R_c \\
2^{-1.5}\rho_c/100 , & r \geq R_c
\end{array}
\right.,
\end{equation}
where $r$ is the distance from the cloud center and $\rho_c = 6\Sigma_c/[(2^{2.5}-1) R_c]=1.6\times 10^{-18}$ g cm$^{-3}$ is the core density. Thus our density profile consists of a constant density in the inner half of the radius, coupled with a powerlaw falloff in the outer half of the radius. Outside this cloud we place a low density ambient medium with a density that is 100 times smaller than the cloud edge density. We choose this density structure because observations indicate the presence of a roughly $r^{-1.5}$ density gradient on large-scales in star-forming clumps \citep[e.g.][]{caselli95a, beuther02c, beuther05b, beuther06a, mueller02a, sridharan05a}. By choosing a flat inner density profile, however, we minimize tidal forces in the cloud core, thereby ensuring maximum opportunity for fragmentation.

We initialize the cloud velocity with a Gaussian-random velocity field with a power spectrum $P(k)\propto k^{-2}$ and a one-dimensional velocity dispersion $\sigma_c = \sqrt{G M_c/ 2 R_c}=2.9$ km s$^{-1}$. The corresponding virial parameter is $\alpha = 5 \sigma_c^2 G M_c/R_c = 5$, so that the turbulent kinetic energy is larger than the potential energy at time zero. However, we do not include any feedback processes (e.g.\ winds or H~\textsc{ii} regions) capable of driving the turbulence, nor do we have other potential driving mechanisms, such as a turbulent cascade from larger scales or ongoing infall. As a result, the turbulence undergoes a rapid decay, which quickly renders the cloud gravitationally bound.

Throughout the computational domain, we initialize the radiation energy density to that of a blackbody with a temperature $T_r = 10$ K. Thus we have $E = a T_r^4 = 7.56 \times 10^{-11}$ erg cm$^{-3}$. Similarly, we initialize the gas temperature within the cloud ($r<R_c)$ to $T_g = 10$ K. Outside the cloud ($r>R_c$), we set the temperature to $T_g = 1000$ K. Since the density outside the cloud is $1/100$ that of the density at the cloud edge, this ensures thermal pressure balance across the cloud boundary. We also set the Planck and Rosseland opacities of the material with $T_g > 500$ K and $\rho < 2^{-1.5}\rho_c/50$ to zero, to ensure that the host ambient medium does not interact with the radiation field, and is not able to cool.

\subsection{Evolution Equations}

The simulations we present in this paper use the ORION adaptive mesh refinement code. The numerical method is nearly identical to that in our previous papers \citep[e.g.][]{krumholz07a, krumholz09c, krumholz10a, myers11a, cunningham11a}. Here we only summarize the physics, and we refer readers to the numerical method papers referenced in Section \ref{sec:numerics} for a full description of ORION's workings. ORION works by solving the equations of compressible gas dynamics including self-gravity, radiative transfer, and radiating star particles, all on an adaptive grid. In our computational domain, we describe every cell with a vector of conserved quantities $(\rho, \rho\vecv, \rho e, E)$, where $\rho$ is the density, $\rho\vecv$ is the momentum density, $\rho e$ is the total internal plus kinetic gas energy density, and $E$ is the radiation energy density in the rest frame of the computational grid. In addition to gas quantities, we also track an arbitrary number of point mass star particles, each of which is described by a position $\vecx_i$, a momentum $\vecp_i$, and an instantaneous luminosity $L_i$, where the subscript $i$ refers to the particle number.

Given this description of the problem, the full set of evolution equations is
\begin{eqnarray}
\label{masscons}
\frac{\partial}{\partial t}\rho & = & - \nabla\cdot(\rho\vecv) - \sum_i \dot{M}_i W(\vecx-\vecx_i) \\
\frac{\partial}{\partial t}(\rho \vecv) & = & -\nabla\cdot(\rho \vecv\vecv) - \nabla P - \rho \nabla \phi - \lambda \nabla E
\nonumber \\
& & {} - \sum_i \dot{\vecp}_i W(\vecx-\vecx_i) 
\label{momcons}
\\
\frac{\partial}{\partial t}(\rho e) & = & -\nabla \cdot [(\rho e+P)\vecv] - \rho \vecv \cdot \nabla \phi - \kappa_{\rm 0P} \rho (4 \pi B - c E) 
\nonumber \\
& & {} + \lambda\left(2 \frac{\kappa_{\rm 0P}}{\kappa_{\rm 0R}} - 1\right) \vecv \cdot \nabla E - \left(\frac{\rho}{m_p}\right)^2 \Lambda(T_g) 
\nonumber \\
& & {} - \sum_i \dot{\mathcal{E}}_i W(\vecx - \vecx_i) 
\label{econsgas}
\\
\frac{\partial}{\partial t}E & = & \nabla \cdot \left(\frac{c\lambda}{\kappa_{\rm 0R} \rho} \nabla E\right) + \kappa_{\rm 0P} \rho (4 \pi B - c E) 
\nonumber \\
& & {} - \lambda \left(2\frac{\kappa_{\rm 0P}}{\kappa_{\rm 0R}} - 1\right) \vecv\cdot \nabla E - \nabla \cdot \left(\frac{3 - R_2}{2} \vecv E\right)
\nonumber \\
& & {}
 +  \left(\frac{\rho}{m_p}\right)^2 \Lambda(T_g) + \sum_i L_i W(\vecx - \vecx_i)
\label{econsrad} 
\\
\label{starmass}
\frac{d}{dt} M_i &= & \dot{M}_i \\
\label{starpos}
\frac{d}{dt} \vecx_i & = & \frac{\vecp_i}{M_i} \\
\label{starmom}
\frac{d}{dt} \vecp_i & = & -M_i \nabla \phi + \dot{\vecp}_i
\\
\label{poisson}
\nabla^2\phi & = &-4\pi G \left[ \rho + \sum_i M_i \delta(\vecx-\vecx_i)\right].
\end{eqnarray}
Equations (\ref{masscons}), (\ref{momcons}), and (\ref{econsgas}) represent the conservation laws for gas mass, momentum, and energy, including terms describing the exchange of these quantities with star particles and with the radiation field. Equation (\ref{econsrad}) is the corresponding conservation of energy equation for the radiation field. Similarly, equations (\ref{starmass}), (\ref{starpos}), and (\ref{starmom}) are the equations of mass and momentum conservation, and the equation of motion, for the point particles. Finally, equation (\ref{poisson}) is the Poisson equation for the gravitational potential $\phi$.  Note that we compute the gas-radiation exchange terms using the mixed frame formulation \citep{mihalas82a, krumholz07b}, allowing us to write them in a form that is manifestly and exactly energy-conserving. 

In these equations, the terms $\dot{M}_i$, $\dot{p}_i$, and $\dot{\mathcal{E}}_i$ represent the rate at which mass, momentum, and energy accrete from the gas onto the $i$th star, and $L_i$ represents the luminosity of that star. We describe how we compute these quantities in Section \ref{sec:numerics}. The quantities $P$ and $T_g$ are the pressure and gas temperature, respectively. These are related by the equation of state
\begin{equation}
P = \frac{\rho k_B T_g}{\mu m_{\rm H}} = (\gamma-1) \rho \left(e - \frac{v^2}{2}\right),
\end{equation}
where $\mu = 2.33$ is the mean molecular weight for molecular gas of Solar composition and $\gamma$ is the ratio of specific heats. We adopt $\gamma=5/3$, appropriate for gas too cool for hydrogen to be rotationally excited, but this choice is essentially irrelevant because $T_g$ is set almost purely by radiative effects. The quantities $\kappa_{\rm 0P}$ and $\kappa_{\rm 0R}$ are the specific Planck- and Rosseland-mean opacities in the rest frame of the gas, $B = c a_R T_g^4/4\pi$ is the Planck function, and $\lambda$ is the flux limiter, given by
\begin{eqnarray}
\lambda & = & \frac{1}{R} \left(\mbox{coth} R - \frac{1}{R}\right) \\
R & = & \frac{|\nabla E|}{\kappa_{\rm 0R} \rho E} \\
R_2 & = & \lambda + \lambda^2 R^2.
\end{eqnarray}
We compute the opacities as a function of the gas density and temperature using the iron normal, composite aggregates dust model of \citet{semenov03a}.

Finally, $\Lambda(T_g)$ represents the line cooling coefficient. We include this because the turbulence can be strong enough in our cluster so that, at isolated points, gas shock-heats to temperatures above a few thousand K. This exceeds the dust sublimation temperature, so the dust opacity becomes nearly zero in this gas. Instead, the gas in this temperature regime is cooled by line emission, which we cannot easily describe with a simple continuum opacity. In this gas, we transfer energy from the gas thermal reservoir to the radiation field at a rate $(\rho/m_p)^2 \Lambda(T_g)$, where $m_p$ is the proton mass, and the function $\Lambda(T_g)$ is taken from \citet{cunningham06a}. See \citet{cunningham11a} for more details of our line cooling approach.

An important subtlety in our evolution equations, which is worth noting, is that we {\it do not} differentiate between gas and dust grain temperatures. At low densities, gas-grain coupling can be imperfect, and it can be important to calculate the two temperatures separately, and to simulate the thermal exchange between dust and gas \citep[e.g.][]{urban09a}. However, grains and gas become very tightly coupled in temperature once the density exceeds $n\sim 10^4 - 10^5$ cm$^{-3}$. For comparison, the mean density in our initial clouds is $n = \overline{\rho}/\mu = 4.0\times 10^5$ cm$^{-3}$. Thus our entire computation is in the strong coupling regime, and there is no need to treat dust and gas temperatures separately.

For simulation ISO, our isothermal run, we modify these equations as follows. First, we omit equation (\ref{econsrad}) entirely. Second we set to zero all terms proportional to $E$ or $\Lambda(T_g)$ in equations (\ref{momcons}) and (\ref{econsgas}). Third, instead of $\gamma=5/3$, we use $\gamma = 1.0001$. This corresponds to neglecting the effects of radiative transfer, and simply keeping the gas temperature almost completely fixed to its initial value.

\subsection{Numerical Method}
\label{sec:numerics}

The ORION code solves equations (\ref{masscons}) -- (\ref{poisson}) in a series of operator-split steps. In each time step, we first integrate the hydrodynamic equations (\ref{masscons}) -- (\ref{econsgas}), excluding the terms describing stars and the radiation field. This update uses a conservative Godunov scheme with an approximate Riemann solver, and is second-order accurate in time and space \citep{truelove98a, klein99a}. Next we solve the Poisson equation (\ref{poisson}) using a multigrid iteration scheme \citep{truelove98a, klein99a, fisher02a}. Third, in the runs where we include radiation, we update the radiation energy equation (\ref{econsrad}) and the radiation terms in the hydrodynamic equations (\ref{masscons}) -- (\ref{poisson}). This update uses the \citet{krumholz07b} conservative update scheme, in which we handle the dominant terms implicitly and the non-dominant terms explicitly. The update for the implicit terms uses the \citet{shestakov08a} pseudo-transient continuation scheme. Finally, we update the stellar quantities, equations (\ref{starmass}) -- (\ref{starmom}), and update gas quantities for the gas-star exchange terms on the right hand sides of equations (\ref{masscons}) -- (\ref{poisson}). We determine the accretion rates of mass, momentum, and energy onto each star by fitting the flow within a radius of four finest-level cells of each star particle to a Bondi-Hoyle flow, following the procedure described by \citet{krumholz04a}. We update the luminosity $L_i$ of each star using the protostellar evolution model described in the appendices of \citet{offner09a}.

Each of these update modules operates within the overall adaptive mesh framework of ORION \citep{berger84a, berger89a, bell94a}. In this scheme, we discretize the computational domain onto a series of levels $l=0,1,2,\ldots, L$. The coarsest level, level 0, has cells of linear size $\Delta x_0$, and covers the entire computational domain. All subsequent levels, with cells of size $\Delta x_l = \Delta x_0/2^l$, cover subregions of the computational domain. Each level consists of a union of rectangular grids of cells, and grids on different levels are nested such that every level $l$ grid with $l>0$ is fully contained within one ore more level $l-1$ grids. To advance a level $l$ in time, we first advance all the grids on that level through a time step $\Delta t_l$, then advance grids on level $l+1$ by two timesteps of size $\Delta t_{l+1} = \Delta t_l / 2$. After the two level $l+1$ advances, we synchronize the boundaries between levels $l$ and $l+1$ to ensure exact conservation of mass, momentum, and energy across level boundaries. The entire update procedure is recursive, so a single advance on level $l+1$ entails two advances of size $\Delta t_{l+2} = \Delta t_{l+1}/2$ on level $l+2$, and so forth to the finest level present. The coarse level timestep $\Delta t_0$ is set by computing the Courant condition on each level (including a contribution to the signal speed from radiation pressure -- \citealt{krumholz07b}) and setting $\Delta t_0 = \min(2^l \Delta t_l)$.

\subsection{Boundary, Refinement, and Star Particle Conditions}
\label{sec:conditions}

At the edge of the computational domain, we use reflecting boundary conditions for the hydrodynamics. However, this choice is irrelevant to the evolution, because our computational domain is large enough to ensure that no material from the cloud ever approaches it. For the gravity, we adopt Dirichlet boundary conditions, with the potential at the computational domain boundary set equal to a multipole expansion of the potential due to the matter in the domain interior, including terms up to the quadrupole. Finally, for radiation we adopt Marshak boundary conditions, with the flux into the computational domain set equal to the flux of an isotropic 10 K blackbody: $F_{\rm in} = c a T_r^4/4 = 0.57$ erg cm$^{-2}$ s$^{-1}$. The boundary condition is equivalent to allowing any radiation generated within the computational domain to escape freely, but also to bathing the computational domain in a 10 K blackbody radiation field.

In order to determine when AMR levels are added or removed, we must also specify refinement conditions. The conditions we use in our simulations are as follows. First, we refine any cell with a density greater than half the edge density of the initial cloud to at least level 1. This ensures that our initial cloud is well resolved. Second, we refine any cell on level $l$ that is within a distance $16\Delta x_l$ or less from any star particle. This ensures that the region around each star is resolved by at least 32 cells on all levels by the finest one. Third, we refine any cell where the density exceeds the local Jeans density \citep{truelove97a},
\begin{equation}
\label{eq:jeans}
\rho > \rho_J = J^2 \frac{\pi c_s^2}{G \Delta x_l^2},
\end{equation}
where $c_s = \sqrt{k_B T_g/\mu}$ is the sound speed. We use a Jeans number $J = 1/8$. Finally, we refine any cell where the local radiation energy gradient satisfied the condition $|\Delta E|/E > 0.15/\Delta x_l$. This ensures that gradients of radiation energy density are always well-resolved. If any of these conditions are met, we refine that point in the computational domain to a higher AMR level, up to the maximum level $L$ for that simulation (see Table \ref{runs}).

Finally, we create a new star in any cell on the maximum level $L$ that violates the Jeans condition, equation (\ref{eq:jeans}), using a Jeans number $J=1/4$. In contrast to previous runs, where we merged star particles together if they approached closer than a certain limit, here we do not allow any star particle that has a mass greater than $0.05$ $\msun$ to be destroyed by merging. Our motivation for choosing this mass limit is that it is roughly the mass at which second collapse to stellar densities occurs \citep[e.g.][]{masunaga98a, masunaga00a}. Objects of lower mass remain extended gas balls with physical sizes of a few AU, and thus are much more likely to merge than the much smaller, more compact protostars they become once they complete their collapse. Complete suppression of mergers for more massive objects is probably an extreme assumption, but as we will see in discussion our results, allowing mergers would only strengthen our conclusions, by moving the stellar mass distribution to higher values.

\section{Simulation Results}
\label{sec:results}

For convenience, throughout this section we will report our results in terms of mean-density free-fall times, where the mean density is $\overline{\rho} = 3 M_c/(4 \pi R_c^3)= 9.4\times 10^{-19}$ g cm$^{-3}$ and the corresponding free-fall time is $t_{\rm ff} = \sqrt{3\pi/32 G \overline{\rho}} = 68.6$ kyr. The free-fall time in the high-density initial core is $\sim 30\%$ shorter, $t_{\rm ff,c} = 52.3$ kyr. In reporting stellar quantities, we only count as stars those star particles with masses above $0.05$ $\msun$, the mass at which second collapse to stellar dimensions occurs. However, this has little effect on our results, since objects below this mass never constitute more than a tiny fraction of the total mass in star particles.

We ran these simulations on a combination of the supercomputers Pleiades at the NASA Advanced Supercomputing facility and Ranger at the Texas Advanced Computing Center. Runs LR, HR, and ISO required roughly 200,000, 850,000, and 60,000 CPU hours, respectively, and ran on between 256 and 960 CPUs (32 to 120 nodes), with the number of CPUs used increasing as a run progressed and the number of high resolution grids increased.

\subsection{Large-Scale Evolution}

\begin{figure}
\plotone{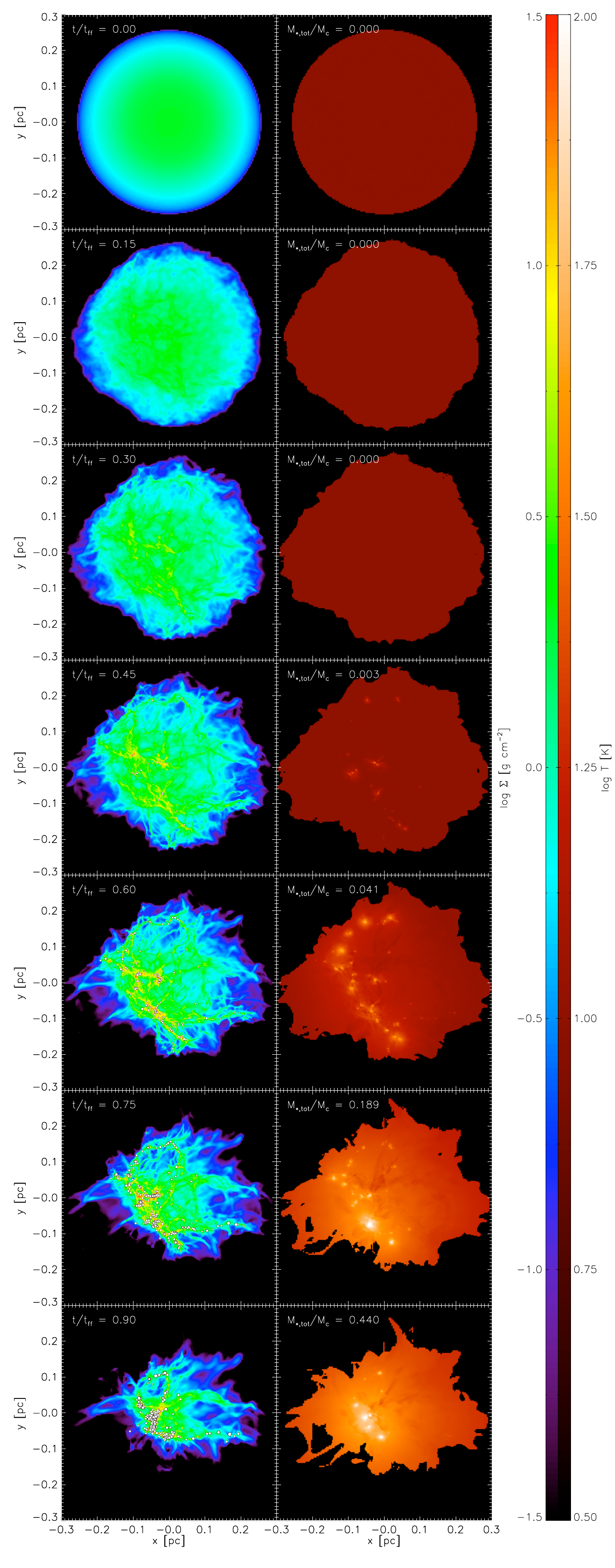}
\caption{
\label{evol_lr}
Column density (left) and density-weighted mean temperature (right) in run LR. The left and right columns show the state of the simulation at the same time, with the time running from $t/t_{\rm ff} = 0$ at the top to $t/t_{\rm ff} = 0.9$ at the bottom, at intervals of $0.15$. The ratio of stellar mass to initial cloud mass, $M_{*,\rm tot}/M_c$, is also indicated in each panel. In the column density plot, white circles show the locations of star particles, with the size of the circle indicating the mass of the star. In the right column, the temperature we show is the radiation temperature, defined by $E = a_R T_r^4$. We show this quantity because the radiation and gas temperatures are nearly identical everywhere except in the low-density, zero-opacity ambient medium, where the radiation temperature is far lower than the gas temperature. By using the radiation rather than the gas temperature, we ensure that our projected temperatures are not artificially enhanced by contributions from the hot ambient medium.
}
\end{figure}

\begin{figure}
\plotone{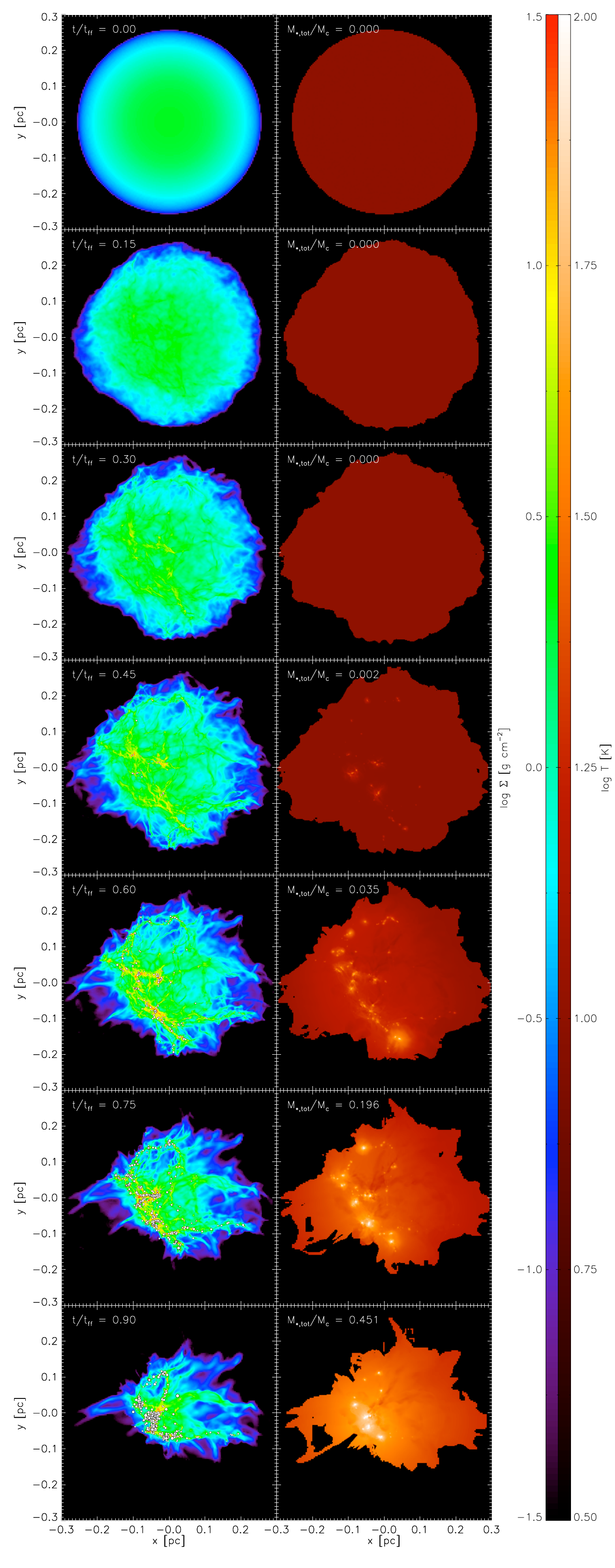}
\caption{
\label{evol_hr}
Same as Figure \ref{evol_lr}, but for run HR.
}
\end{figure}

\begin{figure}
\epsscale{0.7}
\plotone{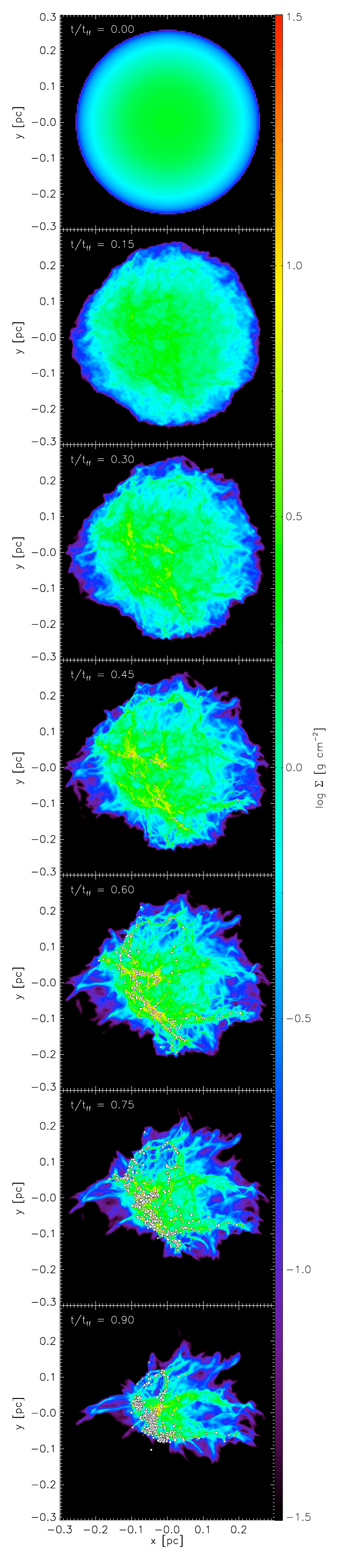}
\epsscale{1.0}
\caption{
\label{evol_iso}
Same as Figure \ref{evol_lr}, but for run ISO. Since this run is isothermal, we show only column density, not density-weighted mean temperature.
}
\end{figure}

Figures \ref{evol_lr}, \ref{evol_hr}, and \ref{evol_iso} show the large-scale evolution of the cloud as it collapses in our simulations. As the plots make clear, the overall distribution of the gas and stellar mass, and the gas temperature structure, is very similar in all runs. As we will see in more quantitative detail later, the evolution of the two radiative runs is very similar in almost every respect, so that we may have confidence that the behavior we are seeing is physical and not a result of resolution effects. Even at late times, the only noticeable difference is the exact positions of individual stars on the periphery of the cloud. These differ primarily because the N-body interactions that occur late in the simulation are chaotic. They can therefore be changed significantly because the amount of gravitational softening in the gas-particle and particle-particle interactions is resolution-dependent.

In both radiative runs, we see that, for the first $(0.3-0.4)t_{\rm ff}$, the initial velocity perturbations we have injected are developing and creating structure, but that no stars have yet formed. The gas temperature remains locked at 10 K, the value imposed by the radiation field. Around $t/t_{\rm ff} = 0.45$, the first stars start to appear at the densest peaks created by the turbulent compression. The mass in stars is still tiny, well under 1\% of the gas mass, and the stars themselves are all quite small. Nonetheless, the effects on the temperature are immediately apparent. Each star is surrounded by a clear region of gas at elevated temperatures. These regions are localized, so that the the bulk of the gas remains cold, and the heated regions around different stars are, for the most part, non-overlapping.

It is not surprising that the formation of stars has such a strong effect. As pointed out by \citet{offner09a}, the energy budget of a star-forming cloud is dominated by the gravitational potential energy released by star formation, even when those stars constitute a tiny fraction of the total mass. This continues to be true up until the point when massive stars with short Kelvin times begin to dominate the bolometric output of the stellar population. In our simulations, even though we do produce $\sim 20$ $\msun$ stars with significant internal luminosities toward the end of the simulations, accretion luminosity is the dominant energy source over most of the simulation time.

This morphology of small regions of warm gas strung out along filaments continues to hold to some extent even at time $t/t_{\rm ff} = 0.6$, when the stellar mass has increased to a few percent of the gas mass. We can still identify distinct heated regions associated with individual stars or small stellar groups, and the bulk of the mass remains near 10 K. In the last two time slices, however, as a larger and larger fraction of the cloud mass is converted into stars, this ceases to be true. Even the coldest gas anywhere in the cloud is now at temperatures noticeably larger than the original background temperature, and the regions of very warm gas, $T\gtsim 100$ K, are beginning to overlap and merge. In the last time slice, the coldest gas anywhere in the computational domain is at $\sim 30$ K, and much of the mass is concentrated in a few compact regions where the temperature is significantly higher. Rather than a few warm, dense regions around individual stars \citep[cf.][]{offner09a} the bulk of the gas is now concentrated into a smaller number of more massive regions that are heated by the collective effects of large numbers of stars.

\subsection{Star Formation History and IMF}

\begin{figure}
\plotone{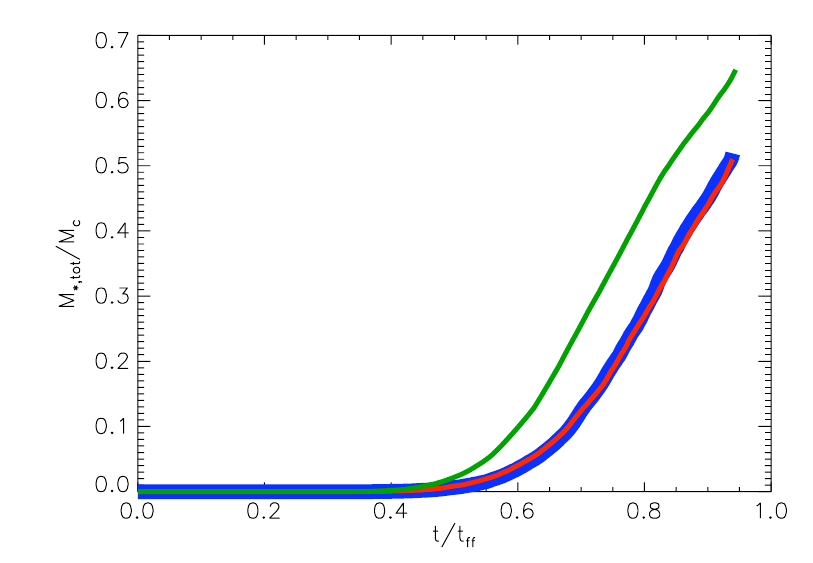}
\caption{
\label{starhist1}
Total mass in stars $M_{*,\rm tot}$, normalized to the initial cloud mass $M_c$, as a function of time, $t$, normalized to the mean-density free-fall time $t_{\rm ff}$. We show results for run LR (thin red line), run HR (thick blue line), and run ISO (thin green line).
}
\end{figure}

\begin{figure}
\plotone{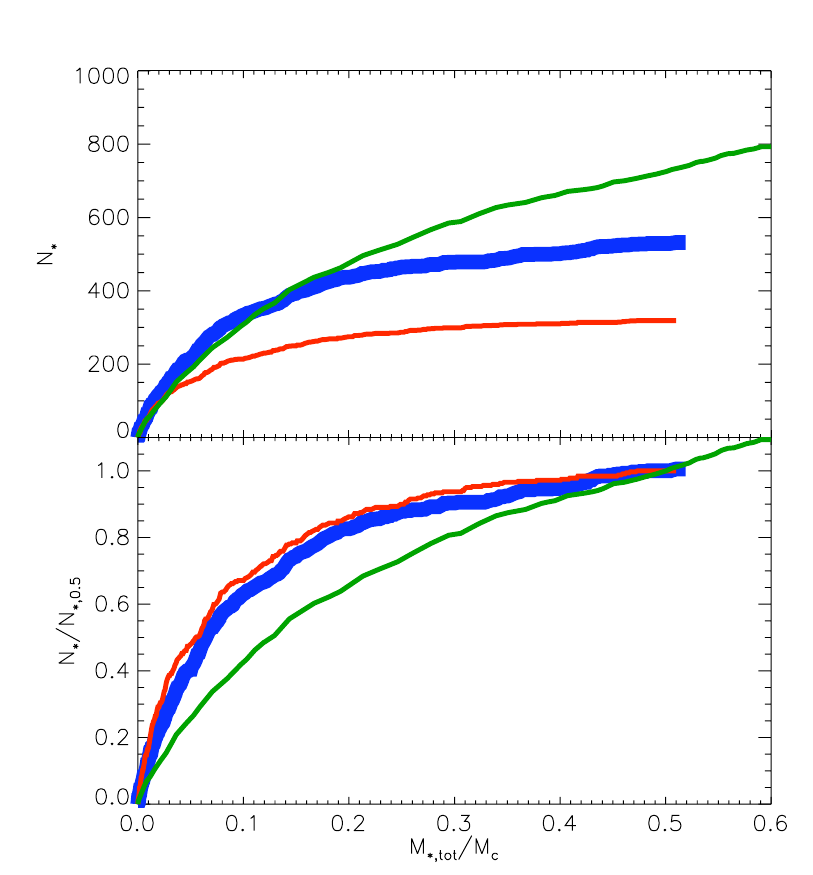}
\caption{
\label{starhist2}
Total number of stars $N_*$ (top) and number of stars normalized to the number present at the time when $M_{*,\rm tot}/M_c = 0.5$ (bottom). We show results for run LR (thin red line), run HR (thick blue line), and run ISO (thin green line).
}
\end{figure}

Figure \ref{starhist1} shows the total mass of all stars as a function of time in the runs. Examining the figure shows that the total mass in stars is nearly identical in the two radiative runs, indicating that this aspect of the simulations is very well converged. 
Run ISO begins to form stars somewhat earlier, and the mass in stars present at equal times is somewhat higher. However, this difference mostly appears to be a time offset. The overall shape in Figure \ref{starhist1} is the same, indicating a generally similar star formation history. The time offset is likely a result of the faster collapse that occurs in the isothermal run, where cooling is assumed to be infinitely rapid and efficient, compared to the radiative run.

Figure \ref{starhist2} shows the number of stars as a function of the total stellar mass in each simulation. The total number of stars is somewhat larger in run HR than in run LR, which is not surprising given the increased resolution. Observations indicate that the binary period distribution is extremely broad, covering separations from only a few stellar radii to $\ga 10^4$ AU \citep{duquennoy91a}. It is therefore not surprising that some binaries that might be resolved into two separate stars in run HR instead appear as a single star in run LR -- indeed, we would expect this result in essentially any simulation that did not resolve the radii of individual stars. Nonetheless, notice that, if we normalize to the number of stars present at equal times and fractions of mass accreted, then the difference between the two runs disappears. The number of stars present at any given time in run HR is roughly $1.6$ times the number present at the same time in run LR. Thus the trend in terms of when the stars are formed in the simulations is nearly identical in the two cases, and we can regard as well-resolved the distribution in time of when stars form. 

The trend of number of stars versus mass shown in Figure \ref{starhist2} is interesting. In the radiative runs, when $M_{*,\rm tot}/M_c \la 0.1$, the number of stars increases roughly linearly with the total stellar mass, as we might expect if the mass per star were constant. However, the rate at which new stars appears drops sharply once $M_{*,\rm tot}/M_c \ga 0.2$. Indeed, we see that $60-70\%$ of all stars have formed at a time when only $\sim 10\%$ of the cloud mass has been incorporated into stars, By the time 20\% of the cloud mass has gone into stars, nearly 90\% of all the stars are in place. In effect, the fragmentation of the gas into new stars has completely shut down. Given that this effect occurs nearly identically in runs LR and HR, this cannot be a resolution effect. In contrast, run ISO shows very different behavior. The number of stars as a function of total stellar mass is almost the same as in run HR up to the point where $\sim 15\%$ of the mass has been incorporated into stars, but the two runs diverge after that. New stars continue forming all the way through run ISO, at a rate that is only slightly less after $M_{*,\rm tot}/M_c \ga 0.2$ than it was earlier in the simulation. This strongly suggests that the shutdown in new star formation we observe in runs LR and HR is a radiative effect, a topic to which we will return in Section \ref{sec:thermo}.

\begin{figure}
\plotone{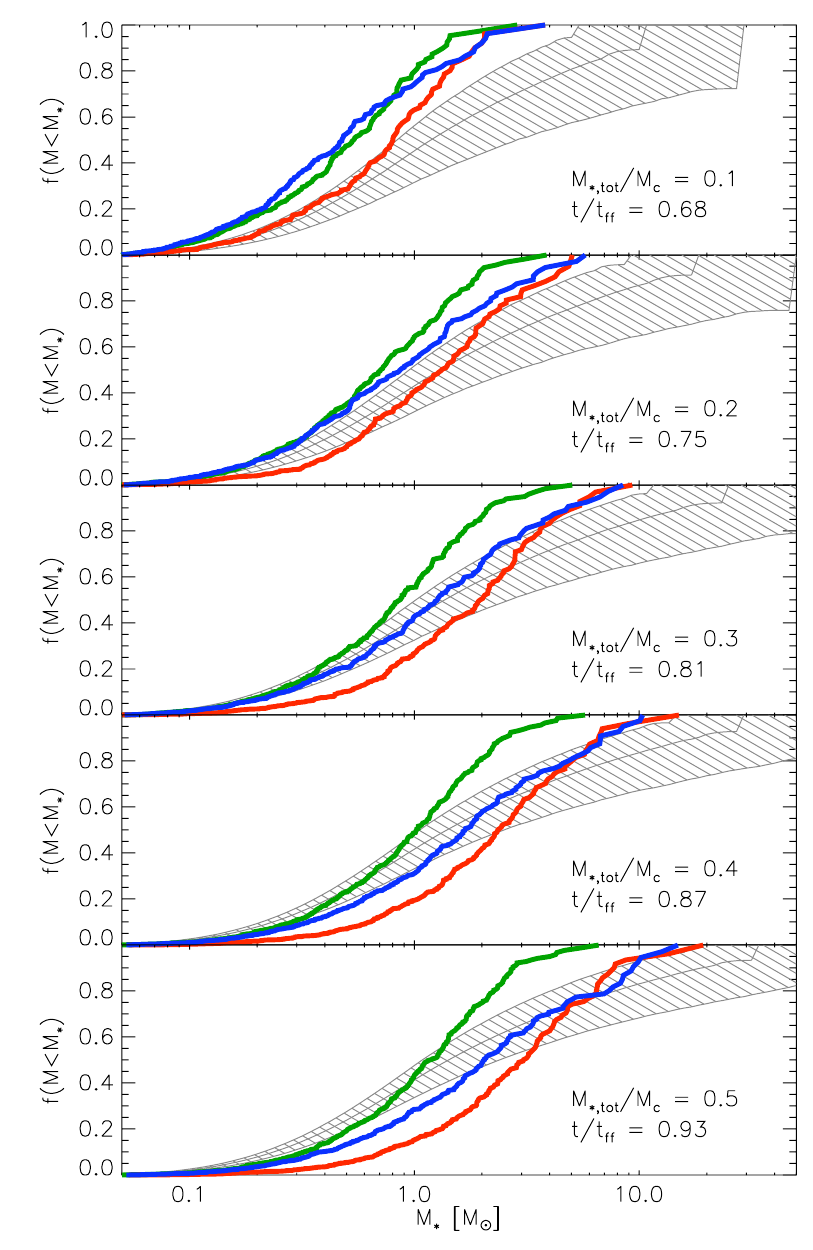}
\caption{
\label{imfplot1}
Fraction $f(<M_*)$ of the total stellar mass in stars with masses smaller than $M_*$, as a function of $M_*$. The five panels show this cumulative mass function in the simulations at the times when the total stellar mass is 10\%, 20\%, 30\%, 40\%, and 50\% of the initial cloud mass as indicated. We also show the times at which simulations LR and HR (but not run ISO) reach these stellar masses, which are identical to within a few percent in runs LR and HR. We show run LR (red line), run HR (blue line), run ISO (green line) and the results of creating a cluster of stars of equal total mass (gray), with each star randomly drawn from a \citet{chabrier05a} IMF following the procedure outlined in Appendix \ref{app:imfsample}. For the \citet{chabrier05a} IMF, the three lines show the 10th, 50th, and 90th percentile of the random drawings, and the hatched region indicates the range between the 10th and 90th percentiles.
}
\end{figure}

\begin{figure}
\plotone{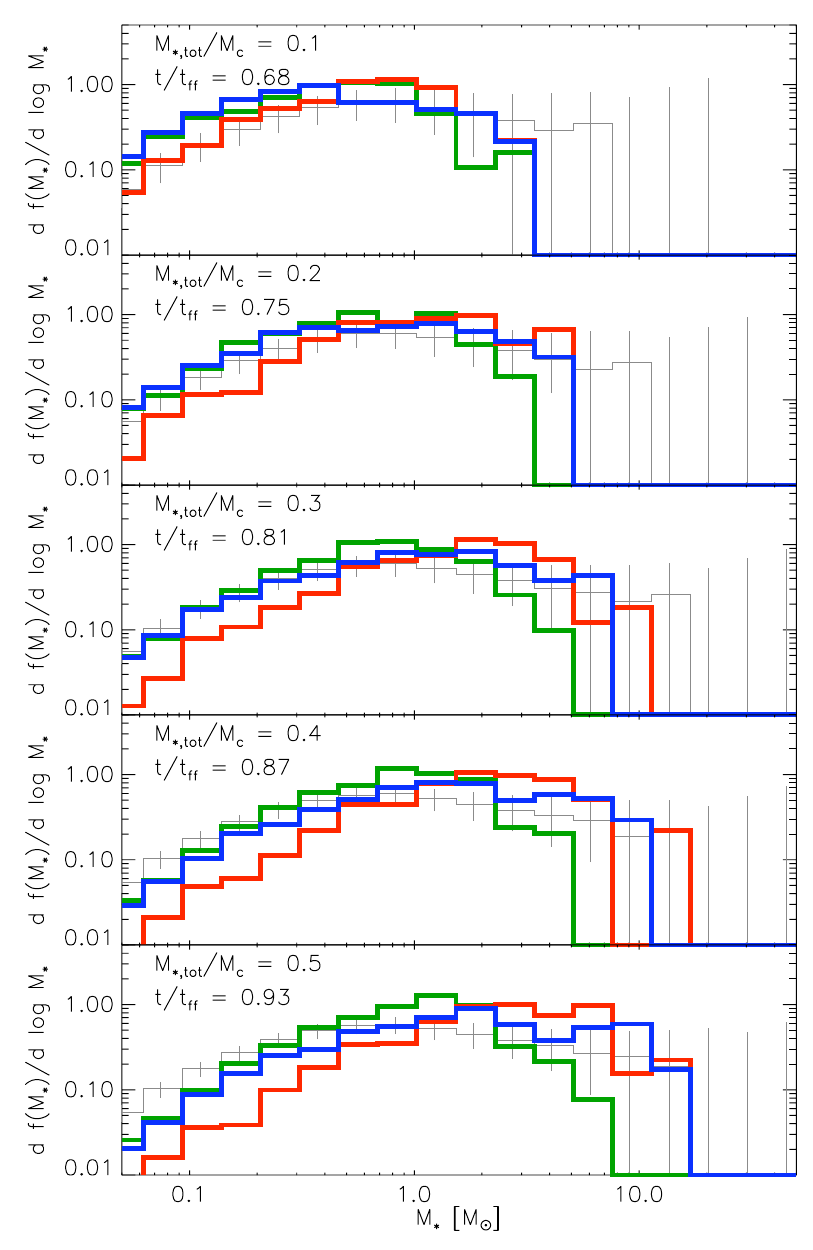}
\caption{
\label{imfplot2}
Same as Figure \ref{imfplot1}, except that we show the differential mass distribution $df(M_*)/d\log M_*$, i.e.\ the value in each bin indicates the fraction of all stellar mass that lies in that bin. We normalized our distributions so that the sum of all bins multiplied by the bin width equals unity. As in Figure \ref{imfplot1}, we show run LR (red line), run HR (blue line), run ISO (green line), and clusters of equal mass drawn from a \citet{chabrier05a} IMF following the procedure in Appendix \ref{app:imfsample} (gray). The histogram values we plot are the 50th percentile of our experiments, and the error bars indicates the range from the 10th to the 90th percentile.
}
\end{figure}

As one might expect, this shutoff of fragmentation into new stars in runs LR and HR even as the total stellar mass continues to increase produces a dramatic effect on the stellar mass distribution. Figures \ref{imfplot1} and \ref{imfplot2} show the cumulative and differential mass distributions of the stars formed in our simulations at the times when the total mass in stars is $10-50\%$ of the initial cluster mass. All these plots show that the stellar mass distribution in the radiative runs moves continuously to higher masses as the simulation proceeds. This is because mass is accreting onto existing stars, which rise in mass, but very few new, lower-mass stars are forming. Note that, while the mean stellar masses are slightly different in runs LR and HR, the systematic drift of these mean to higher masses as the total stellar mass rises appears to about occur equally in both runs. In run ISO, on the other hand, there is much less evolution in the shape of the IMF. The fraction of mass in very small objects does decrease slightly with time, but the IMF in run ISO peaks at $\sim 1$ $\msun$ in every time slice. Quantitatively, we find that, from the point where $M_{\rm *,tot}/M_c \approx 0.15$ and the star formation histories in runs ISO and HR begin to diverge, up to the point when $M_{\rm *,tot}/M_c \approx 0.5$ and run HR ends, the mass-weighted median stellar mass\footnote{Defined as the mass $m$ for which stars with masses $m_*<m$ comprise half the total stellar mass} in run ISO increases by only a quarter of a dex, while in run HR it increases by half a dex. Thus the behavior of run ISO is similar to that in previous simulations done with prescribed equations of state\footnote{We cannot directly compare to the earlier radiative simulations of low mass clusters by \citet{bate09a} and \citet{offner09a}, because these produced fewer than 20 objects. Their IMFs are therefore much too sparsely sampled for it to be possible to make any meaningful statements about their time-dependence.} (e.g.~see Figure 1 of \citet{bonnell04a}, which a similar increase in median mass from $0.7-1.0$ free-fall times in their simulation.)\footnote{There may also be difference between our simulations and those of \citeauthor{bonnell04a}\ due to differences in initial conditions (partly centrally condensed for us versus uniform density for them) and equation of state (isothermal for our run ISO, barotropic for them.)}

For comparison, we have generated 10,000 clusters each of mass 100, 200, 300, 400, and 500 $\msun$, randomly drawn from a \citet{chabrier05a} IMF,\footnote{The argument has been advanced in the literature that the observed IMF in clusters with masses as small as a few hundred $\msun$ is is truncated at high masses compared to Chabrier or similar IMF (\citet{weidner10a}, but see \citet{lamb10a}, \citealt{calzetti10a}, and \citet{fumagalli11a} for observational counterarguments). Here we are mostly interested in the peak of the IMF, not the high mass end where our simulations have too few stars to make statistically strong statements. We therefore proceed with the simplest assumption that there is no high mass truncation, since it makes no difference for our purposes.} with a minimum mass of $0.05$ $\msun$ and a maximum of $150$ $\msun$. We properly account for finite sampling using the procedure described in Appendix \ref{app:imfsample}. As the plots show, the mass distribution of stars formed in the radiative simulations drifts to systematically higher masses than the observed IMF once $\sim 30-50\%$ of the mass has been turned into stars. The disagreement is highly significant, and occurs at stellar masses that are extremely well-resolved in the simulations. For example, consider Figure \ref{imfplot2} at the time when $M_{*,\rm tot}/M_c = 0.5$. For run HR at that time, the mass in almost every bin from $1-10$ $\msun$ is above the 90th percentile of random drawings from a Chabrier IMF, while the mass in almost every bin below $1$ $\msun$ is below the 10th percentile of random drawings from a Chabrier IMF. Indeed, a Kolmogorov-Smirnov comparison between the mass functions produced in the simulations and the Chabrier IMF shows that, with the exception of the HR run at the point when $M_{*,\rm tot}/M_c = 0.3$, all the mass functions shown in Figures \ref{imfplot1} and \ref{imfplot2} are inconsistent with having been drawn from the Chabrier IMF at confidence levels better than 1 part in $10^6$.

In contrast, run ISO is consistent with the IMF at the low mass end at essentially all times. It is deficient in massive stars compared to a Chabrier IMF, an effect that has been observed before in simulations without radiative transfer \citep{maschberger10a} and taken as evidence for the so-called ``IGIMF (integrated galactic initial mass function) effect". We obtain the same result here, but find that it disappears in simulations that include radiation.

One might be tempted to fix this problem simply by scaling all the stellar masses by some factor less than unity, to account for mass ejected by protostellar outflows, which we have not included. However, because the peak of the IMF is evolving with time in our simulations, a scaling factor that produces agreement between the simulated IMF and the observed one at one time would not produce agreement at earlier or later times. The central problem is not so much that the IMF in the simulation is too top-heavy, but that the median mass increases continuously with time. However, it does seem likely that protostellar outflows can help solve the problem by reducing the star formation rate and thus the luminosity, as we discuss further in Section \ref{sec:solutions}. Such an effect cannot be captured by a simple rescaling of the masses, though.

\subsection{Gas Thermodynamics and Fragmentation}
\label{sec:thermo}

\begin{figure*}
\plotone{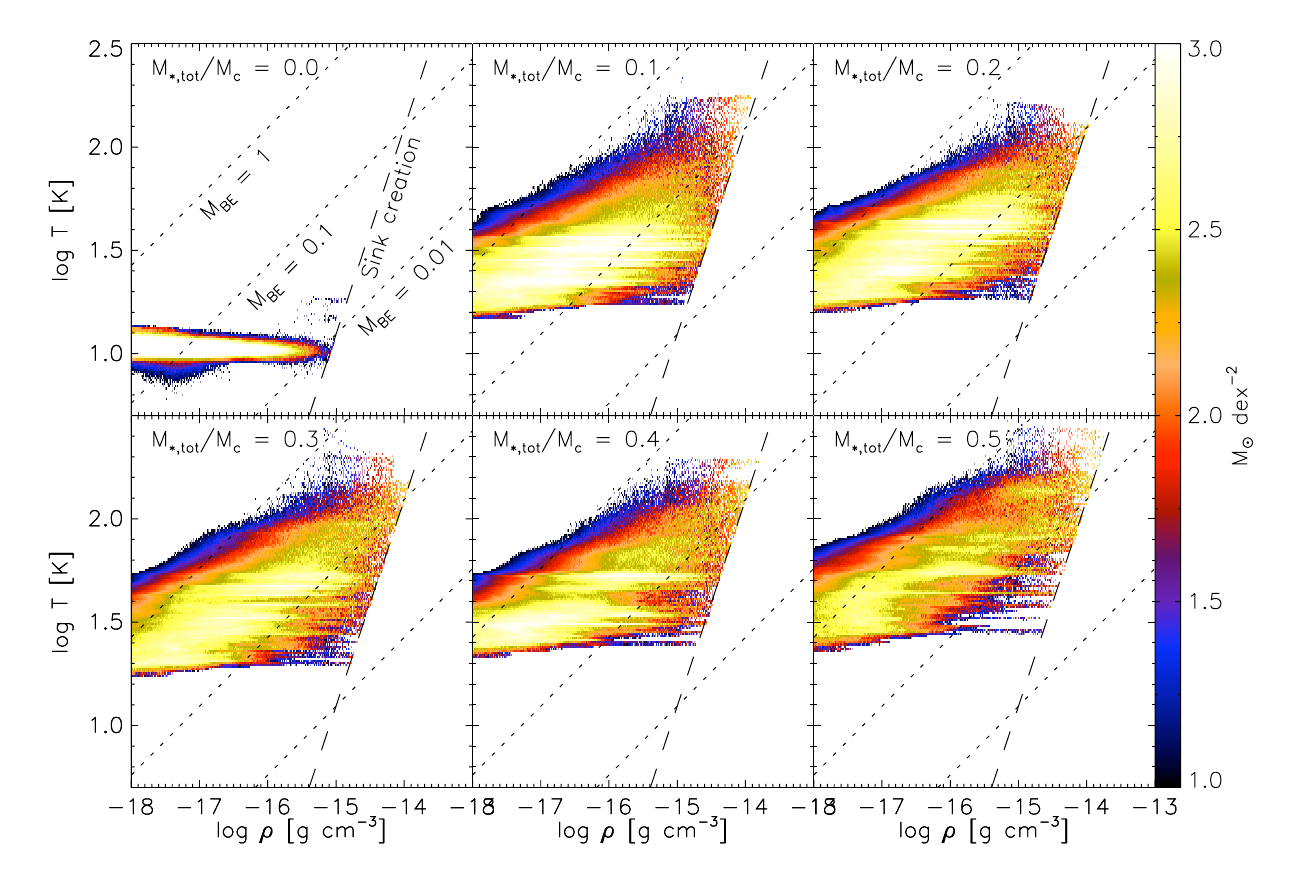}
\caption{
\label{phaseplot_lr}
Distribution of the gas mass in bins of density and temperature in run LR. The panels show the distributions at the time when the total stellar mass normalized to the initial cloud mass is $M_{*,\rm tot}/M_c = 0.0$, 0.1, 0.2, 0.3, 0.4, and 0.5, as indicated. For the time $M_{*,\rm tot}/M_c = 0.0$ we select the last time snapshot for which $M_{*,\rm tot} = 0$. Within a panel, the color of each pixel indicates the density of mass in that pixel, measured in $\msun$ per dex in density per dex in temperature. The sashed black line labelled ``sink creation" indicates the locus of density and temperature for which a computational cell exceeds the sink particle creation condition, equation (\ref{eq:jeans}) evaluated with $J=1/4$. Gas cells that fall to the right of this line create sink particles that incorporate some of their mass, which is why no cells fall to the right of the line. For comparison, the dotted lines indicate the loci where the Bonnor-Ebert mass $M_{\rm BE} = 0.01$, $0.1$, and $1$ $\msun$, as indicated.
}
\end{figure*}

\begin{figure*}
\plotone{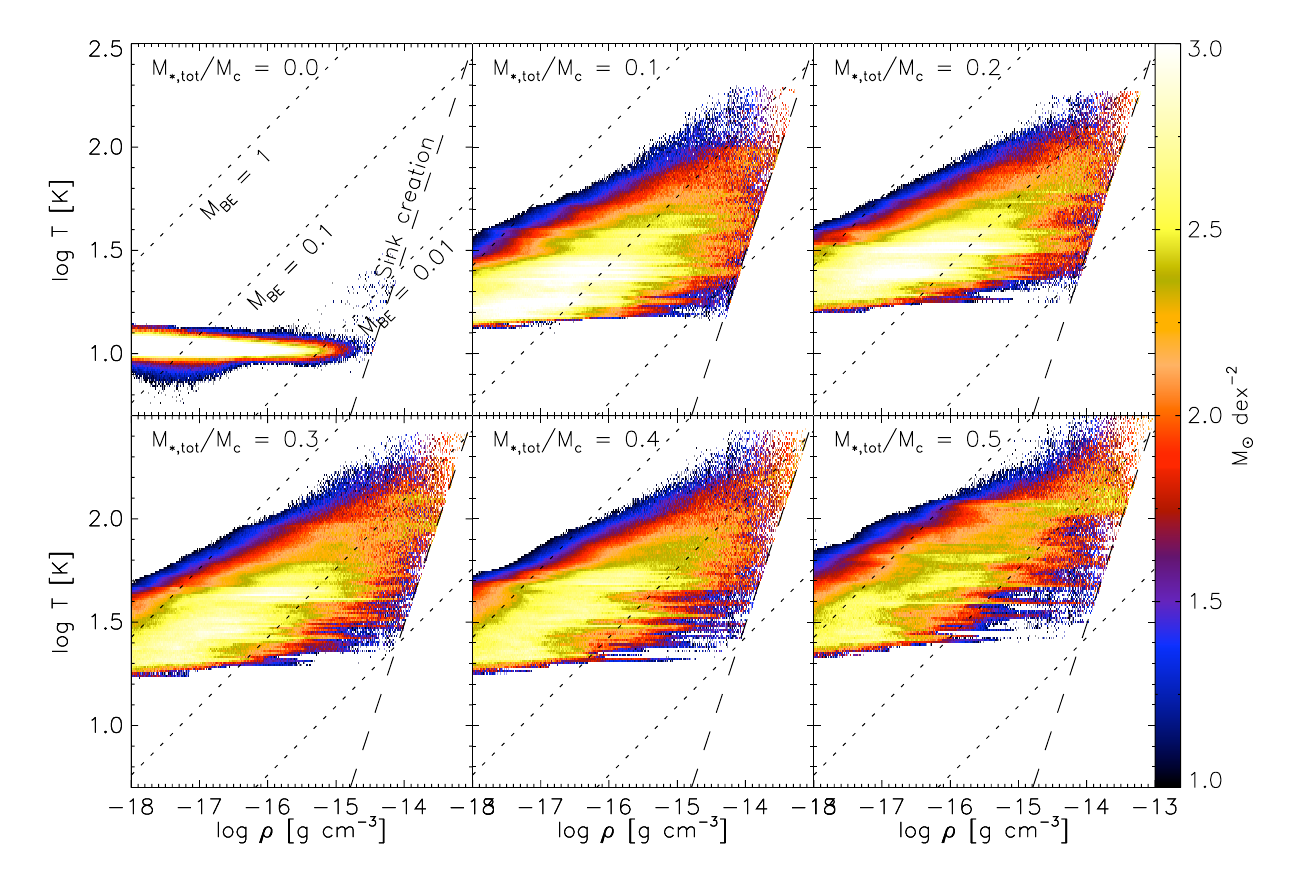}
\caption{
\label{phaseplot_hr}
Same as Figure \ref{phaseplot_lr}, but for run HR. Note that the sink creation locus is shifted to higher densities by factor of 4 compared to run LR.
}
\end{figure*}

\begin{figure}
\plotone{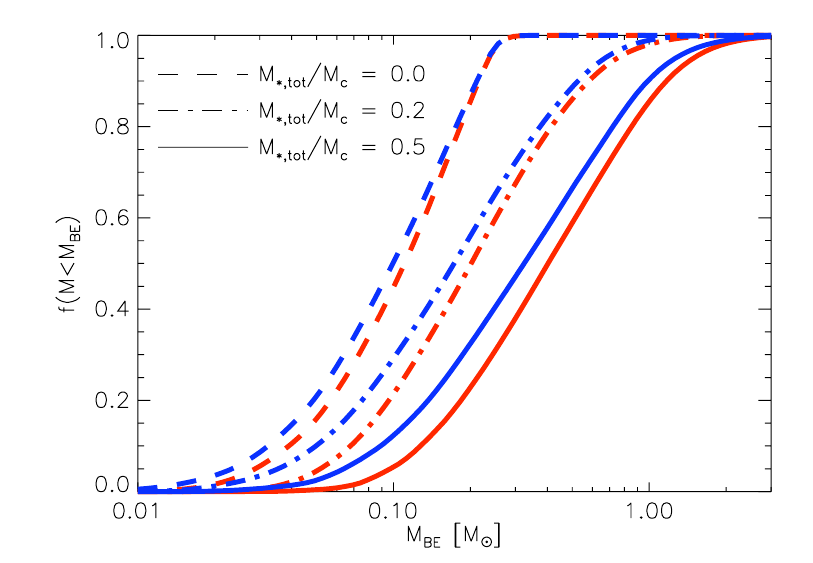}
\caption{
\label{phaseplot_1d}
Fraction $f(M<M_{\rm BE})$ of the gas mass for which the Bonnor-Ebert mass at the local density and temperature, as computed from equation (\ref{eq:mbe}), is less than $M_{\rm BE}$. We show these mass distributions for run LR (red) and run HR (blue), at the times when the total mass is $M_{*,\rm tot}/M_c = 0.0$ (dashed, just before the first star forms), $0.2$ (dot-dashed), and $0.5$ (solid).
}
\end{figure}

The reason for the shutdown in fragmentation and the drift to systematically higher stellar masses with time in the radiative runs becomes clear if we consider how the gas density and temperature evolve with time. Figures \ref{phaseplot_lr} and \ref{phaseplot_hr} show the distribution of gas mass in the density - temperature plane as star formation proceeds in runs LR and HR. For comparison, we also overlay lines of constant Bonnor-Ebert mass, where
\begin{equation}
\label{eq:mbe}
M_{\rm BE} = 1.18 \frac{c_s^3}{\sqrt{G^3 \rho}},
\end{equation}
and $c_s = \sqrt{k_B T/\mu}$ is the isothermal sound speed. The Bonnor-Ebert mass is significant because objects with masses below $M_{\rm BE}$ can be supported against collapse by thermal pressure. We therefore expect that the lowest mass stars to form will tend to have masses comparable to the smallest values of $M_{\rm BE}$ found in the gas. Even if turbulence does create fragments with masses below $M_{\rm BE}$, these will be stable against collapse as a result of their thermal pressure. Figure \ref{phaseplot_1d} summarizes this result by showing how the gas mass is distributed with respect to $M_{\rm BE}$ at different times in the simulation. As the plot shows the runs are not completely converged at the low $M_{\rm BE}$ end, but the general trend that the mean Bonnor-Ebert mass systematically increases is clear in both runs.

Figures \ref{phaseplot_lr} -- \ref{phaseplot_1d} show that, immediately before any stars have formed, the great majority of the mass has a temperature within a few K of 10 K, the initial temperature and the temperature imposed by the background radiation field. Consequently, the densest material in the cloud is in a density and temperature regime where $M_{\rm BE} \sim 0.01$ $\msun$, and objects of this mass are able to collapse. Nearly half the mass in the cloud lies in the region between $M_{\rm BE} = 0.01$ $\msun$ and $0.1$ $\msun$, so there is plenty of material available to make low mass stars. Once stars begin to form, however, their radiation raises the temperature significantly, pushing to higher $M_{\rm BE}$. This increase is partly offset by an increase in the mean density, but the density does not increase quickly enough to compensate for the rapidly rising temperature -- likely because the density rise occurs on a timescale associated with the mean density free-fall time, while the temperature rise is driven by stellar accretion occurring at the peaks of the density distribution, which operate on a much shorter timescale. As a result of this evolution, there is not much material that is dense and cold enough to make small stars. For example, if we look run at HR, we find that $20\%$ of the gas mass has $M_{\rm BE} < 0.05$ $\msun$ just before the first star forms, and thus is able to make the smallest stars we consider. In contrast, the mass fraction able to create such small stars drops to 10\% at the point when $M_{*,\rm tot}/M_c = 0.2$, and to only 2\% when $M_{*,\rm tot}/M_c = 0.5$. Thus we see that the formation of new stars has stopped because it is no longer possible for small fragments to gravitationally collapse. By the end of the run, the smallest gravitationally-unstable fragments are approaching 1 $\msun$ in size.

The underlying physical reason for this effect, of course, is the radiation released by the already-formed stars. This in turn, is primarily driven by accretion luminosity, with a subdominant contribution from nuclear burning and Kelvin-Helmholtz contraction later in the simulation.

\section{Discussion}
\label{sec:discussion}

\subsection{The Overheating Problem}

A systematic increase in the mean stellar mass induced by heating of the gas due to accretion luminosity is a new phenomenon in simulations of star cluster formation. Radiative suppression of fragmentation has been reported in the literature before, but no previous simulation has observed it to shift the typical stellar mass scale in regions as large as entire star clusters. We emphasize that, even if we regard the absolute stellar mass peak we obtained as uncertain due to resolution effects, the trend of increasing mean mass is robust, and appears equally strong in both simulations. It has not been seen in previous work due to the limitations we outlined in Section \ref{sec:intro}. Most simulations of large-scale cluster formation with initial conditions similar to ours, which are typical of Galactic star formation, have not included radiative transfer. They have adopted a simple equation of state, which puts in by hand the result that the peak stellar mass is invariant \citep[e.g.][]{bonnell04a}. We effectively recover the same results in our run ISO: the median stellar mass does increase slightly with time, but the increase is significantly smaller than in the radiative runs, and is consistent with what has been found in earlier non-radiative simulations.

Those simulations that have included radiation have either focused on regions too small or with too few stars \citep[e.g.][]{bate09a, offner09a, offner10a} to produce the overlap of the heating regions around many stars we observe, or have focused on single massive cores, where suppression of fragmentation is expected \citep[e.g.][]{krumholz07a, krumholz10a, myers11a}. Indeed, in these contexts, radiative suppression of fragmentation is necessary to obtain agreement between simulations and observations. For single massive cores, suppressed fragmentation has tentatively been seen in high resolution interferometric observations \citep{bontemps10a, longmore11a}. In the absence of radiation, the disks formed in simulations of low mass star formation tend to undergo excessive fragmentation, leading to an overproduction of brown dwarfs relative to stars and to various other conflicts with observation \citep{luhman07a}. Including radiation fixes this problem \citep{bate09a, offner09a, offner10a}. Indeed, \citet{bate09a} argues that the observed peak of the IMF can be explained as arising from the mass scale at which radiative feedback halts fragmentation. While this argument is plausible, it relies on the assumption that we can consider the bubble of radiatively-warmed gas around each star to be isolated from other stars, amidst a background of cool gas. This assumption holds in the low-mass, low-density regions simulated by \citet{bate09a} and \citet{offner09a,offner10a}, where regions of heating are $\sim 0.05$ pc in size, much smaller than the interstellar separation. It clearly does not hold in our simulation, both because our stars are closer together than in a low mass star-forming region, and because our heating regions are larger due to the higher accretion rates produced by the higher gas densities.  This suggests that the critical problem in our simulation is that the regions of warm gas around individual stars begin to overlap. As a result, all the gas in the cluster is heated, rather than simply discrete regions.

One might hope to avoid this problem by halting star formation early on, before enough mass goes into stars to allow the heated regions to overlap. However, such a solution seems to require improbable fine tuning. Examining figures \ref{evol_lr} and \ref{evol_hr}, we see that the overlap of hot regions is well underway by the time 30-40\% of the mass has been incorporated into stars. Figures \ref{imfplot1}, and \ref{imfplot2} show that the shift of the simulation IMF to higher masses than the \citet{chabrier05a} IMF is also largely complete by this point. Since this is about the minimum star formation efficiency required to have any possibility of making bound clusters \citep{kroupa01b, fall10a}, the fact that at least some star formation does result in bound clusters suggests that the star formation efficiency cannot be vastly lower than this value most of the time.

\subsection{Understanding the Problem}

We can estimate the dividing line between the two cases of heating in discrete regions around single stars and heating in the bulk of the protocluster gas using the analytic radiative transfer approximation of \citet{chakrabarti05a, chakrabarti08a}, coupled to the formalism developed by \citet{krumholz08a}. \citeauthor{chakrabarti05a} consider a spherical cloud of dusty gas with radius $R$, mass $M$, and a density profile $\rho\propto r^{-k_\rho}$, surrounding a point source of radiation of luminosity $L$, with dust whose specific opacity depends on wavelength as $\kappa=\kappa_0 (\lambda_0/\lambda)^{\beta}$. In such a cloud, they show that the temperature profile approximately follows
\begin{equation}
\label{eq:tprofile}
T = T_{\rm ch} \left(\frac{r}{R_{\rm ch}}\right)^{-k_T},
\end{equation}
where $r$ is the distance from the cloud center, $R_{\rm ch}$ and $T_{\rm ch}$ are the characteristic radius and temperature of the dust photosphere formed within the cloud, and $k_T$ is a powerlaw index to be approximated by a numerical fit. For convenience we define $\Sigma = M/\pi R^2$, $\eta = L/M$ (measured in cgs units, not Solar units), $\alpha = 1/[2\beta+4(k_\rho-1)]$, $\tilde{R} = R/R_{\rm ch}$, and $T_0 = hc/k_B \lambda_0$. For Milky Way dust, $\beta\approx 2$ and $\kappa_0 \approx 0.54$ cm$^2$ g$^{-1}$ at $\lambda_0 = 100$ $\mu$m \citep{weingartner01a}, but the results depend on these parameters very weakly. With these definitions, the characteristic radius and temperature and the powerlaw index are given by
\begin{eqnarray}
\frac{R_{\rm ch}}{R} & = & \left\{ \left(\frac{\eta}{4\sigma_{\rm SB} \tilde{L}}\right)^{\beta} \Sigma^{4+\beta} \left[\frac{3-k_\rho)\kappa_0}{4(k_\rho-1)T_0^\beta}\right]^4\right\}^\alpha \\
T_{\rm ch} & = & \left\{ \left(\frac{\eta}{4\sigma_{\rm SB} \tilde{L}}\right)^{k_\rho-1} \Sigma^{k_\rho-3}
\left[\frac{4(k_\rho-1) T_0^\beta}{(3-k_\rho)\kappa_0}\right]^2\right\}^\alpha \\
\tilde{L} & \approx & 1.6 \tilde{R}^{0.1} \\
k_T & \approx & \frac{0.48 k_\rho^{0.05}}{\tilde{R}^{0.02k_\rho^{1.09}}} + \frac{0.1 k_{\rho}^{5.5}}{\tilde{R}^{0.7k_\rho^{1.9}}}.
\label{eq:kt}
\end{eqnarray}
The latter two expressions are approximations based on fits to numerical solutions of the radiative transfer equation, and reproduce the numerical results with high accuracy.

\begin{figure}
\plotone{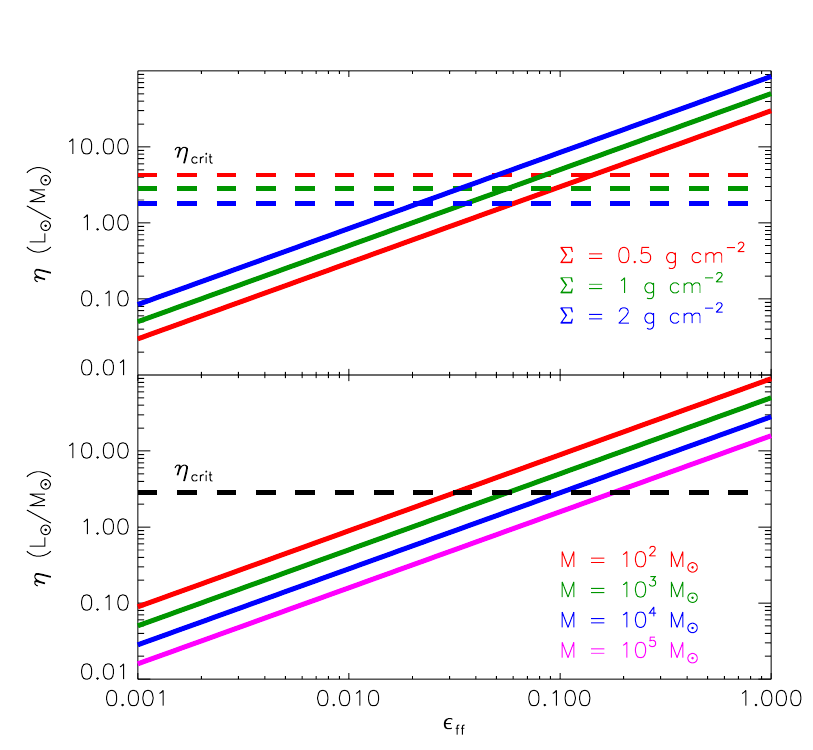}
\caption{
\label{fig:fragsuppress}
Luminosity to mass ratio $\eta$ versus dimensionless star formation rate $\epsilon_{\rm ff}$. In each panel, dashed horizontal lines show the critical light to mass ratio $\eta_{\rm crit}$ at which heating regions around individual protostars merge and fragmentation is suppressed. Solid lines show the value of $\eta_{\rm acc}$ for an accretion-powered cluster-forming gas cloud with the indicated value of $\epsilon_{\rm ff}$. In the upper panel we show clouds with $\Sigma = 0.5$, $1$, and $2$ g cm$^{-1}$ (red, green, and blue), all of mass $M = 10^3$ $\msun$. In the lower panel we show clouds of fixed $\Sigma=1$ g cm$^{-2}$, with masses $M = 10^2$, $10^3$, $10^4$, and $10^5$ $\msun$ (red, green, blue, and purple).
}
\end{figure}

A rough condition for the heating regions around individual protostars to merge and heat the bulk of the gas is that the combined luminosity $L$ of all the protostars, which we approximate as being near the cloud center, be high enough so that the temperature $T$ at the edge of the cloud, $r = R$, be higher than the background temperature $T_b\approx 10$ K to which gas settles when it is not heated by a nearby star. Thus, to avoid overheating we require that the luminosity to mass ratio $\eta$ be smaller than the value $\eta_{\rm crit}$ for which $T(R) = T_b$. For a given cloud mass $M$ and surface density $\Sigma$, it is straightforward to use equations (\ref{eq:tprofile}) -- (\ref{eq:kt}) to numerically determine the value $\eta_{\rm crit}$ for which the condition $T(R) = T_b$ is satisfied. In what follows we do so for a background temperature $T_b = 10$ K and density profile $k_\rho = 3/2$, roughly what is seen in massive star-forming clumps \citep[e.g.][]{mueller02a}, but the result we obtain is not very sensitive to this choice.

The luminosity is in turn related to the star formation rate in the simulations. \citet{krumholz08a} show that accretion onto low mass stars yields an energy release per unit mass accreted $\psi \approx 2.1\times 10^{14}$ erg g$^{-1}$. This number is a result of stellar structure considerations, which fix the characteristic radii of protostars. Thus the light to mass ratio $\eta_{\rm acc}$ in a cloud of mass $M$ powered by accretion luminosity from stars forming at a rate $\dot{M}_*$ is
\begin{equation}
\eta_{\rm acc} = \psi \frac{\dot{M}_*}{M} = \epsilon_{\rm ff} \frac{\psi}{t_{\rm ff}},
\end{equation}
where $t_{\rm ff}=\sqrt{\pi R^3/8 G M}$ is the mean-density free-fall time of the cloud and $\epsilon_{\rm ff}$ is the dimensionless star formation rate introduced by \citet{krumholz05c}.

In Figure \ref{fig:fragsuppress}, we plot $\eta_{\rm crit}$ and $\eta$ for clouds of varying mass $M$ and surface density $\Sigma$ as a function of $\epsilon_{\rm ff}$. Our values of $M$ and $\Sigma$ are chosen to span the range of typical cluster-forming gas clumps in the Milky Way. The value of $\eta_{\rm crit}$ of course depends on $\Sigma$ alone, while that of $\eta_{\rm acc}$ is proportional to $\epsilon_{\rm ff}$. The plot shows that, for any plausible cloud mass and surface density, if $\epsilon_{\rm ff}\ga 0.1$ then $\eta > \eta_{\rm crit}$. This plot explains why, in our simulation, the stellar heating regions overlap. In our simulation, $\sim 50\%$ of the gas is in stars when $t/t_{\rm ff} \approx 1$, so we have $\epsilon_{\rm ff} \approx 0.5$. This puts us in the regime in which heating zones overlap, and fragmentation is suppressed. In contrast, the simulations of \citet{bate09a}, \citet{offner09a}, and \citet{peters10b} have substantially lower surface densities, $\Sigma \la 0.1$ g cm$^{-2}$, putting them in the regime where heating zones do not overlap and fragmentation is unlikely to be suppressed even with quite high $\epsilon_{\rm ff}$, except within the disks around each star. The simulations of \citeauthor{offner09a}, since they include driven turbulence, also have a lower value of $\epsilon_{\rm ff}$.

Note that this argument is consistent with the one made by \citet{elmegreen08a} for why the Jeans mass should vary little in regions where the dust and gas  temperatures are well-coupled, like those we study. The crux of their argument is that increases in the gas density lower the Jeans mass, but also produce a higher star formation rate, which in turn raises the dust temperature and the Jeans mass. The two effects nearly offset one another. However, this offset only occurs if the star formation rate and the density are related by a volumetric \citet{schmidt59a, schmidt63a} law with $\epsilon_{\rm ff} \approx 0.01$ (see their equation 6). If $\epsilon_{\rm ff}$ rises with time, as it does in our simulation, then the Jeans mass will not remain independent of density.

\subsection{Possible Solutions to the Problem}
\label{sec:solutions}

Understanding the problem also suggests an immediate solution. \citet{krumholz07e} compile observations from the literature and find that, even in dense, cluster-forming gas clumps, $\epsilon_{\rm ff} \sim 0.01$. \citet{evans09a} obtain similar values of $\epsilon_{\rm ff}$ in cluster-forming regions observed by the c2d Survey, although c2d targets clusters considerably more diffuse and lower in mass than the one we have simulated here. Figure \ref{fig:fragsuppress} shows that such clouds are not in the regime where heating zones overlap and fragmentation is suppressed, unless their masses are quite low, $M \la 10^2$ $\msun$. This explains why fragmentation is not suppressed in real clouds.

However, this result has important implications for simulations of star cluster formation. It implies that, once radiation physics is included in the simulations, one cannot expect to obtain the correct IMF without also obtaining the correct star formation rate, or at least a star formation rate that is roughly correct. In simulations that do not include radiation feedback, no such care is needed. Fortunately, obtaining the correct star formation rate in simulations is not terribly difficult. Simulations where the turbulence is driven artificially can reproduce the observed value $\epsilon_{\rm ff} \approx 0.01$. Even better, simulations that include stellar wind feedback naturally produce realistic, low values of $\epsilon_{\rm ff}$ without any artificial driving \citep[e.g.][]{li06b, nakamura07a, wang10a}, and preliminary evidence indicates that this does reduce accretion luminosities to the point where fragmentation is suppressed far less than we have found (Hansen et al., 2011, in preparation). It is not clear if this effect is scalable to all cluster masses \citep{fall10a}, but it does suggest a way toward simulations of cluster formation that simultaneously obtain the correct star formation rate and the correct IMF.

It is thus possible that the problem might be solved the the inclusion of other physics that our simulation omits, such as outflows and photoionization. These mechanisms might be able to generate regions of high enough density that their Bonnor-Ebert masses will be low even in the presence of overlapping regions of radiative heating.

\subsection{Implications for Competitive Accretion versus Core Accretion}

It is also interesting to consider the implications of our result for the competitive accretion versus core accretion models for the formation of star clusters and origin of the IMF. Roughly speaking, the competitive accretion model is that collapses that produce star clusters are global in nature, so all stars accrete from the same mass reservoir, and the stellar mass distribution is determined by a competition between formation of new, small fragments (which pushes the mean mass to lower values) and growth of existing fragments by Bondi-Hoyle accretion (which pushes the mean mass to higher values) \citep[e.g.][]{bonnell01a, bate05a, bonnell06c}. In contrast, in the core accretion model, collapses that produce individual star systems are local rather than global, so that different protostars are for the most part not accreting from the same mass reservoir. In this case, the mass distribution of the stars is set by the mass distribution of the regions of localized collapse, the ``cores" \citep[e.g.][]{padoan02a, mckee03a, padoan07a, alves07a, hennebelle08b, hennebelle09a}. Intermediate models are also possible, in which low mass stars form via local collapse, but either massive stars or the cores from which they grow form via a global collapse \citep[e.g.][]{peretto06a, wang10a}.

\citet{krumholz05a} point out, and \citet{bonnell06c} and \citet{offner08b} confirm, that which mode of star formation takes place depends on the level of turbulence and on $\epsilon_{\rm ff}$. If the turbulence is sub-virial, or becomes sub-virial through decay that is not offset by internal feedback or external driving, then $\epsilon_{\rm ff}$ becomes large and competitive accretion is the dominant star formation mode; core accretion prevails if the turbulence remains at virial levels and $\epsilon_{\rm ff}$ is small. In our simulation we do not include either artificial driving or any physical feedback mechanisms capable of driving the turbulence (e.g.\ protostellar outflows or H~\textsc{ii} regions), so our simulation produces large $\epsilon_{\rm ff}$ and we obtain a competitive accretion-like mode of star formation. However, crucially, we have shown that such a mode of star formation cannot produce the correct IMF due to the radiative suppression problem we have identified. The constant production of new, low-mass stars on which competitive accretion relies to keep accretion onto existing stars from pushing the IMF to ever-increasing masses does not happen once radiative feedback is included, at least in the minimal case where hydrodynamics, gravity, and radiative feedback are the only physical ingredients. It is conceivable that some mechanism we have omitted might still enable the production of low mass stars even in clusters with high $\epsilon_{\rm ff}$ (e.g.\ fragmentation induced by expanding H~\textsc{ii} shells), but in this case that mechanism would be responsible for controlling the peak of the IMF. Our results therefore suggest the minimal competitive accretion model is not compatible with the observed IMF.

One might try to alleviate this problem by choosing significantly less dense initial conditions and retaining the high $\epsilon_{\rm ff}$ required for competitive accretion, i.e.\ by selecting a lower $\Sigma$ in Figure \ref{fig:fragsuppress}. However, this solution faces a severe problem: the initial conditions we have selected are typical of the observed gaseous properties of clouds where massive star formation occurs \citep[e.g.][]{shirley03a, faundez04a, fontani05a}. Surface densities are even larger in globular clusters, yet these show the same IMF peak as the field \citet{de-marchi00a, de-marchi10a}. If the minimal competitive accretion model can only reproduce the observed IMF from initial conditions far less dense than we have simulated, then its applicability is limited to low-density regions like Taurus, which generally do not contain any massive stars.

\subsection{Implications for Fragmentation-Induced Starvation}

\begin{figure}
\plotone{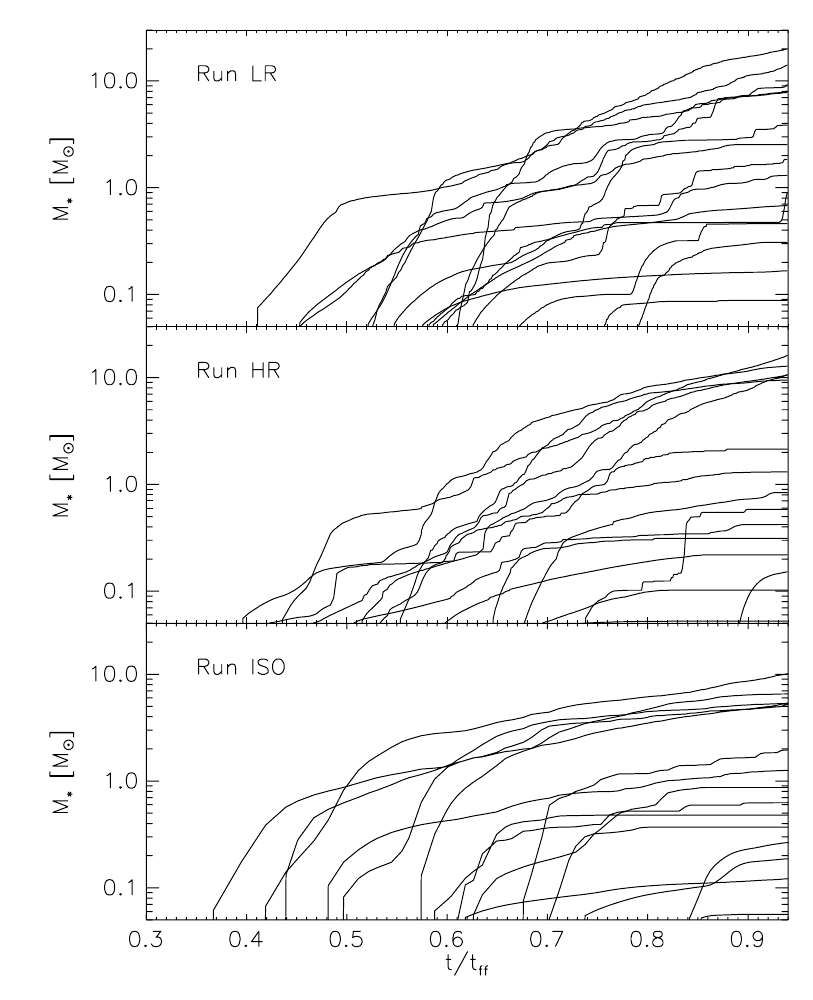}
\caption{
\label{mdot}
Stellar mass as a function of time for a sample of individual stars in runs LR (top), HR (middle), and ISO (bottom). The stars shown are the five most massive at the final time in the simulations, plus 10 other stars evenly distributed in mass.
}
\end{figure}

\begin{figure}
\plotone{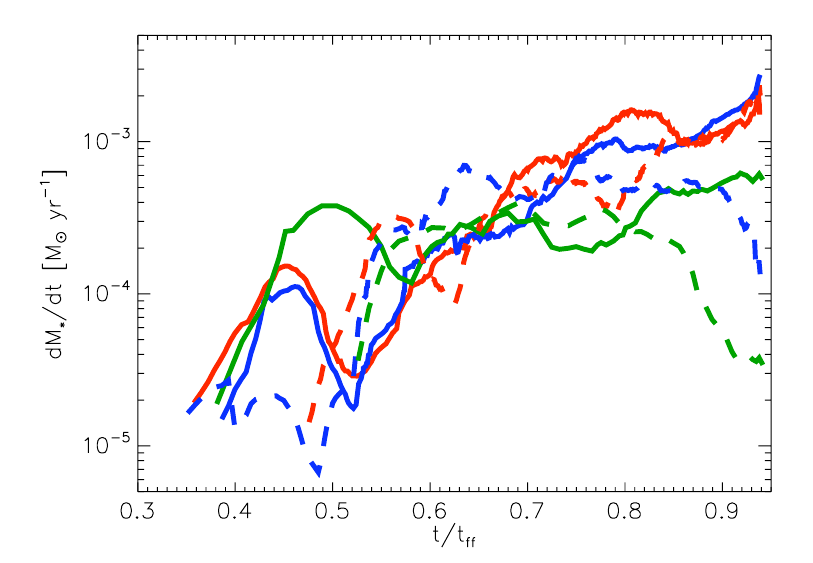}
\caption{
\label{starvation}
Mass accretion rate for the most massive (solid) and second most massive (dashed) stars at the final time in runs LR (red), HR (blue), and ISO (green). To minimize confusion, the accretion rates have been smoothed over a timescale of $0.05 t_{\rm ff}$.
}
\end{figure}

It is also interesting to examine how individual stars, and particularly the most massive stars, grow in mass. We show this in Figures \ref{mdot} and \ref{starvation}, which show the mass versus time and the mass accretion rate versus time for a sample of stars in each run. In run LR the most massive star we form is 20.0 $\msun$, in run HR the most massive star is 16.2 $\msun$, and in run ISO it is 10.3 $\msun$. In runs LR and HR, the most massive stars are continuing to grow rapidly at the end of the simulation, with accretion rates that are generally flat or increasing with time. The most massive stars are also growing in run ISO, but more slowly and with accretion rates that are either constant or declining with time.

These results have potential implications for the idea of fragmentation-induced starvation proposed by \citet{peters10b}. In their simulations (which have a resolution comparable to that of our run LR), the most massive stars stop growing after a certain point because the accretion flow that is feeding them fragments to produce small stars rather than being accreted by the massive star. Thus massive stars exhibit accretion rates that fall with time. We do so something roughly consistent with this behavior in run ISO, but not in our radiative runs. This is likely an effect of radiative suppression of fragmentation.

\citet{peters10b} also include radiative transfer in their simulations, but they do not find strong suppression of fragmentation. This is probably because their simulated cloud has a much lower column density ($\Sigma\approx 0.03$ g cm$^{-2}$) than either our simulated clouds or than typical regions of massive star formation in the Galaxy ($\Sigma\sim 1$ g cm$^{-2}$). \citet{krumholz10a} show that the amount by which radiation suppresses fragmentation is highly sensitive to the column density, and predict essentially no suppression at the column density used by \citeauthor{peters10b} The physical reason for this is that a cloud with $\Sigma=0.03$ g cm$^{-2}$ is optically thin even in the near-infrared, so starlight that is absorbed by dust grains promptly escapes, and most gas is not heated by the radiation. It is therefore not surprising that \citeauthor{peters10b}~see fragmentation-induced starvation and we do not.

We emphasize, however, that the absence of fragmentation-induced starvation in our radiative runs does not mean that fragmentation-induced starvation does not occur under typical Galactic star-forming conditions. We have just argued that fragmentation is suppressed too strongly in our simulations because star formation is too rapid. Indeed, simulations indicate that outflows allow more fragmentation to occur even in single massive cores than in comparable simulations without outflows \citep{cunningham11a}. However, our results suggest that, before fragmentation-induced starvation can be considered an important mechanism in regulating massive star formation, it will be necessary to simulate the formation of a star cluster using typical Galactic conditions, like we do, and to include mechanisms that produce realistically low star formation rates.

\section{Summary}
\label{sec:summary}

We report simulations of the formation of a massive star cluster comparable in size to the Orion Nebula Cluster. Our simulations use adaptive mesh refinement to obtain high resolution, and include radiation-hydrodynamics coupled to a realistic treatment of stellar radiative feedback. These are the first simulations reported in the literature that include radiation feedback in the context of the typical region of Galactic star cluster formation, as opposed to focusing on single low-mass \citep{commercon10a} or high-mass \citep{krumholz07a, krumholz10a, myers11a} cores, or on low-mass or low-density regions like Taurus \citep{bate09a, offner10a, peters10b}.

Our simulations return a surprising result. At early times in the simulations, accreting stars produce bubbles of warm, radiatively-heated gas around themselves, and within these bubbles fragmentation is suppressed by the increased Bonnor-Ebert mass. However, we find that, once $\sim 10-20\%$ of the gas in the protocluster has been converted to stars, these bubbles of warm gas begin to overlap and merge. Rather than resembling a few warm islands surrounded by a sea of cold gas, we instead have a cloud where all the gas is warmed by the collective luminosity of all the accreting stars. 

Once the simulation reaches this state, radiation feedback raises the temperature and the Bonnor-Ebert mass throughout the remaining gas enough to essentially halt the formation of any further stars. Mass continues to be converted from gas to stars, but this is almost entirely through accretion onto existing stars rather than formation of new ones. As a result, when radiation is included, the stellar mass distribution in a globally-collapsing star cluster such as the one we simulate is not nearly constant or very slightly increasing with time, as has been reported in earlier, non-radiative simulations, and as we find here in a control run that does not include radiation. Instead, the stellar mass distribution shifts strongly to systematically higher masses as star formation proceeds, eventually becoming too top-heavy compared to the observed IMF. While the absolute mass scale remains uncertain in our simulations due to our inability to resolve tight binaries, the result that the IMF is non-constant and increasing with time is robust against changes in resolution. This implies that, unless there is also some mechanism to ensure that star formation in every protocluster stops when the IMF peak is in the same place, it is not possible to produce the invariant IMF peak that we observe via the global collapse scenario we have simulated.

We argue that the underlying reason that this problem occurs is that, in the absence of either external turbulent driving or any sort of internal mechanical feedback to slow star formation down, stars in our simulation form too quickly. Since accretion luminosity produced as gas falls onto stars is what ultimately drives the temperature increase in our simulations that shuts off fragmentation and leads to a top-heavy IMF, the problem is likely to be alleviated in simulations that include enough physics to obtain a low star formation rate similar to that observed in real star clusters. We are in the process of conducting such simulations now, and will report on the results in future publications.

\acknowledgements We thank R.~Banerjee, C.~Federrath, R.~Klessen, M.~Mac Low, \& T.~Peters for helpful discussions. This work was supported by an Alfred P.\ Sloan Fellowship (MRK); the NSF through grants CAREER-0955300 (MRK) and AST-0807739 (MRK), and AST-0908553 (CFM and RIK); NASA through ATFP grant NNX09AK31G (RIK, CFM, and MRK) and a Spitzer Space Telescope Theoretical Research Program grant (CFM and MRK); and the US Department of Energy at LLNL under contrast DE-AC52-07NA (RIK). Support for computer simulations was provided by an LRAC grant from the NSF through Teragrid resources and NASA through grants from the ATFP and Spitzer Theory Program.

\begin{appendix}

\section{Generating Comparison IMF Samples}
\label{app:imfsample}

The statistical samples for the IMFs shown in Figures \ref{imfplot1} -- \ref{imfplot2} consist of stellar populations drawn from a \citet{chabrier05a} IMF, subject to the constraint that the total mass of the population have specified value. We create each cluster by the following procedure. First, we draw stars from the \citeauthor{chabrier05a} IMF,
\begin{equation}
\label{eq:imf}
\frac{dn}{d\ln M_*} = \mathcal{N} \left\{
\begin{array}{ll}
\exp(-[\ln \{M_*/\msun\} - \ln 0.2]^2/2\sigma^2), & M_* \leq \msun \\
\exp[-(\ln 0.2)^2/2\sigma^2] (M_*/\msun)^{-1.35}, \quad & M_* > \msun
\end{array}
\right.,
\end{equation}
where $\mathcal{N}$ is a normalization constant and $\sigma = 0.55\ln 10$. We truncate this mass function at $0.05$ $\msun$ on the lower end (to match our minimum stellar mass in the simulation) and at $120$ $\msun$ on the upper end. We continue to draw stars so as long as the total mass of stars is smaller than the specified target mass. If we draw a star of a mass such that adding it to our population causes the total mass to exceed the target mass by more than $0.1$ $\msun$, we reject it and draw another. We continue drawing until the total mass of stars is within $0.1$ $\msun$ of the target mass.

Once we have a set of stars, we form the cumulative and differential distributions. We repeat this procedure 10,000 times each for clusters of total mass (from $100-500$ $\msun$. To produce the values shown in Figures \ref{imfplot1} and \ref{imfplot2}, at each mass $M_*$ on the $x$-axis, we sort the values of the 10,000 cumulative or differential distributions at that value of $M_*$. The 10th, 50th, and 90th percentiles shown are the values at that mass point or mass bin are the 1,000th, 5,000th, and 9,000th vales in the sorted lists.

\end{appendix}

\bibliographystyle{apj}
\bibliography{refs}

\end{document}